\documentclass[11p,a4paper]{article}
\pdfoutput=1
\usepackage{jheppub}
\usepackage[utf8]{inputenc}
\usepackage{latexsym,amsfonts,amsmath,amscd,amssymb,epsf}
\usepackage{hyperref}
\usepackage{soul,xcolor}
\usepackage{natbib}
\usepackage{graphicx}

\usepackage{bm}
\let\vec=\bm

\newcommand{\beq}{\begin{equation}}
\newcommand{\eeq}{\end{equation}}
\newcommand{\cK}{{\cal K}}
\newcommand{\cP}{{\cal P}}

\title{In-medium gluon radiation spectrum with all-order resummation of multiple scatterings in longitudinally evolving media}

\author[a]{Carlota Andres,}
\author[b,c]{Liliana Apolin\'ario,}
\author[d]{Fabio Dominguez}
\author[d]{and Marcos Gonzalez Martinez}
\emailAdd{carlota.andres-casas@polytechnique.edu}
\emailAdd{liliana@lip.pt}
\emailAdd{fabio.dominguez@usc.es}
\emailAdd{marcosg.martinez@usc.es}

\affiliation[a]{CPHT, CNRS, \'Ecole polytechnique,
Institut Polytechnique  de Paris, 91120 Palaiseau, France}
\affiliation[b]{LIP, Av. Prof. Gama Pinto, 2, P-1649-003 Lisboa, Portugal}
\affiliation[c]{Departamento de F\'{\i}sica,  Instituto Superior T\'ecnico, Universidade de Lisboa, Av. Rovisco Pais 1, 1049-001 Lisboa, Portugal}
\affiliation[d]{Instituto Galego de F\'isica de Altas Enerx\'ias IGFAE, Universidade de Santiago de Compostela, Santiago de Compostela, 15782 (Spain)}
\begin{document}

\abstract{Over the past years, there has been a sustained effort to systematically enhance our understanding of medium-induced emissions occurring in the quark-gluon plasma,
driven by the ultimate goal of advancing our comprehension of jet quenching phenomena. To ensure meaningful comparisons between these new calculations and experimental data, it becomes crucial to model the interplay between the radiation process and the evolution of the medium parameters, typically described by a hydrodynamical simulation. This step presents particular challenges when dealing with calculations involving the resummation of multiple scatterings, which have been shown to be necessary for achieving  an accurate description of the in-medium emission process.  In this paper, we extend our numerical calculations of the fully-resummed gluon spectrum to account for longitudinally expanding media. This new implementation allows us to quantitatively assess the accuracy of previously proposed \emph{scaling laws} that establish a correspondence between an expanding medium and a ``static equivalent''. Additionally, we show that such scaling laws yield significantly improved results when the static reference case is replaced by an expanding medium with the temperature following a simple power-law decay. Such correspondence will enable the application of numerical calculations of medium-induced energy loss in realistic evolving media for a broader range of phenomenological studies.}

\maketitle

\section{Introduction}
\label{sec:intro}

Since the beginning of the heavy-ion collision programs at the Relativistic Heavy-Ion Collider (RHIC) and the Large Hadron Collider (LHC), the investigation of jets and high-$p_T$ hadrons has emerged as a pivotal element in comprehending the behavior of the strongly interacting matter produced under such extreme conditions, particularly the quark-gluon plasma (QGP) \cite{Connors:2017ptx,Busza:2018rrf,Cunqueiro:2021wls,Apolinario:2022vzg}. In high-energy nuclear collisions, the initial hard scattering generates highly energetic partons, which subsequently traverse all stages of the hot and dense system's evolution ultimately fragmenting into final state jets. By examining the modifications experienced by these heavy-ion jets compared to proton-proton jets, valuable insights regarding the properties and dynamics of QCD matter can be obtained.

Given that the medium-induced emission of gluons is the primary mechanism responsible for the energy loss of hard partons and the modification of jets, the early focus of the jet quenching community revolved around studying the medium-induced radiation spectrum and  computing energy loss  observables (see e.g. refs.~\cite{Casalderrey-Solana:2007knd,Mehtar-Tani:2013pia,Blaizot:2015lma,Qin:2015srf} for theory reviews and \cite{Zapp:2012ak,Zapp:2013vla,Marquet:2009eq,Armesto:2009zi,Renk:2011gj,Wang:2013cia,Andres:2016iys,Cao:2017hhk,Hulcher:2017cpt,Casalderrey-Solana:2018wrw,He:2018xjv,Huss:2020dwe,Huss:2020whe,Zigic:2021rku,Mehtar-Tani:2021fud,JETSCAPE:2021ehl,Xie:2022ght,JETSCAPE:2022jer,Xie:2022fak,Luo:2023nsi,Mehtar-Tani:2024jtd,Brewer:2018dfs,Caucal:2020xad} for a few phenomenological energy loss studies). While the first calculations of this spectrum were conducted over 20 years ago \cite{Baier:1996kr, Baier:1996sk, Zakharov:1996fv,Zakharov:1997uu,Gyulassy:2000er, Gyulassy:2000fs, Wiedemann:2000za}, recent years have witnessed renewed efforts aimed at enhancing their precision by relaxing several of the commonly employed approximations to derive analytical results. These efforts include relaxing the high-energy approximation to account for the medium transverse dynamics \cite{Sadofyev:2021ohn,Barata:2022krd,Andres:2022ndd,Barata:2023qds}, going beyond the multiple soft (or harmonic oscillator) approximation when resumming multiple scatterings \cite{Zakharov:2004vm,Caron-Huot:2010qjx,Feal:2018sml, Andres:2020vxs,Mehtar-Tani:2019tvy,Barata:2020sav,Andres:2020kfg,Andres:2022bql}, incorporating a proper matching to the infrared sector \cite{Moore:2021jwe,Schlichting:2021idr,Yazdi:2022bru}, or relaxing the soft approximation for the radiated gluon \cite{Blaizot:2012fh,Apolinario:2014csa,Isaksen:2023nlr}. These advancements have significantly contributed to deepen our understanding of the in-medium radiation spectrum, thereby paving the way for more precise phenomenological analyses compared to previous achievements.

One of the main challenges in integrating these recent theoretical advancements directly into phenomenological studies is to accurately consider the dynamic evolution of the plasma, which is typically described through hydrodynamic simulations. In principle, all that is required is a well-founded model that establishes the relationship between the medium parameters used in the Quantum Chromodynamics (QCD) calculations (density of scatterings, Debye mass, jet quenching parameter, etc) and those coming from the hydrodynamic model (temperature, energy density, etc). However, this introduces additional complexities when evaluating the in-medium spectrum, especially in approaches involving the resummation of multiple scatterings to all orders \cite{Zakharov:2004vm,Caron-Huot:2010qjx,Moore:2021jwe,Schlichting:2021idr,Yazdi:2022bru,Feal:2018sml, Andres:2020vxs,Mehtar-Tani:2019tvy,Barata:2020sav,Andres:2020kfg,Andres:2022bql}. Prior to incorporating such realistic conditions, new theoretical developments addressing spectrum calculations are often examined in the simplified case of a ``brick'' --- a constant temperature  plasma with fixed length. This setup introduces additional symmetries which can be exploited to analytically evaluate some of the integrations, particularly those involving the longitudinal direction. However, beyond the static brick scenario, analytical expressions for the spectrum accounting for multiple scatterings have only been derived within the framework of the harmonic oscillator (HO) approximation and for particular functional forms of the jet quenching parameter $\hat q$ dependence on the longitudinal coordinate~\cite{Arnold:2008iy, Adhya:2019qse,Adhya:2021kws,Adhya:2022tcn}. Some of these findings remain applicable in the context of the improved opacity expansion~\cite{Mehtar-Tani:2019tvy,Barata:2020sav}. Nevertheless,  incorporating the medium evolution in the parameters entering the radiation spectrum, beyond these limited scenarios, significantly increases the computational time required for precise evaluations. This issue becomes particularly challenging in phenomenological analyses where one needs to compute the in-medium emission spectrum for a large number of possible trajectories of the initial hard parton.

It would be then highly desirable to be able to pre-compute the emission spectrum for a well-defined set of parameters, enabling the efficient utilization of these pre-tabulated spectra in subsequent phenomenological studies. However, accomplishing this task is not straightforward due to the spectrum's dependence on the parameters' values at each point along the trajectories of the hard partons. As a result, there is no direct method of evaluating an arbitrary trajectory from a finite number of pre-evaluated ones. To tackle this issue, a solution was proposed in~\cite{Salgado:2002cd,Salgado:2003gb}. This approach consists of formulating a set of \emph{scaling laws} that establish a connection between an arbitrary trajectory in the QGP and an ``equivalent static scenario''. These scaling laws were initially proposed as an ansatz and tested against the very limited available analytical results, considering both multiple scatterings within the HO approximation and the single scattering regime given by the first order in opacity \cite{Gyulassy:2000er,Gyulassy:2000fs,Wiedemann:2000za}. It is worth noting that this approach has never been tested against realistic trajectories extracted from hydrodynamical simulations, essentially because at the time there were no numerical implementations of a multiple scattering approach to in-medium radiation for an expanding medium with an arbitrary evolution of the medium parameters.

In this manuscript, we make use of the recent developments in the numerical evaluation of the in-medium radiation spectrum beyond the HO approximation, using a full Yukawa parton-interaction model \cite{Gyulassy:1993hr}, developed in refs.~\cite{Andres:2020vxs,Andres:2020kfg}. Even though this framework was initially developed in the context of brick calculations, it allows for the straightforward inclusion of the medium parameters' evolution, enabling the evaluation of the in-medium spectrum along trajectories extracted from hydrodynamical simulations. The computational intensity of this approach means that it is not practical to compute the emission spectrum for every single trajectory needed in a thorough phenomenological study, but instead we have the possibility of testing the scaling laws formulated in~\cite{Salgado:2003gb} against a numerical evaluation of the emission spectrum in realistic evolving media. We initially conduct a comprehensive analysis with quantitative comparisons between the spectra along the trajectories sampled from the hydrodynamic simulation and their corresponding static scenarios given by the scaling laws ~\cite{Salgado:2003gb}, showing that there are sizeable discrepancies in the low-energy part of the spectrum. As a way of improving the overall accuracy, we propose an alternative method for matching spectra from arbitrary hydrodynamic trajectories to those obtained from profiles that follow a power-law decay. As anticipated, this alternative approach proves to be significantly more accurate to matching against the static case. These findings show that it is possible to have a pre-computed set of spectra for a power-law which can accurately approximate any realistic trajectory, thus paving the way for future phenomenological analyses employing the fully resummed medium-induced spectrum \cite{Andres:2020vxs,Andres:2020kfg}.

The paper is organized as follows: in section~\ref{sec:full} we present the all-order in-medium radiation framework derived in~\cite{Andres:2020vxs,Andres:2020kfg}, focusing on its generalization to longitudinally evolving media. Section~\ref{sec:matching} is dedicated to the analysis of the static and power-law scaling laws. We compare the resulting spectra obtained using these scaling laws with those obtained from trajectories sampled over a smooth-averaged hydrodynamic simulation of the QGP produced in heavy-ion collisions at the LHC. Results on the effectiveness of the proposed scaling laws for trajectories sampled from a hydrodynamic simulation accounting for fluctuations event by event are shown in section~\ref{sec:ebye}. In section~\ref{sec:conclusions}, we summarize the main findings and conclusions of the study. We further provide three appendices. Appendix~\ref{sec:appA} presents additional results for the power-law scaling law using different values of the power and initialization parameters. Appendix~\ref{sec:appKLN} analyzes the performance of the power-law scaling within a smooth-averaged hydrodynamics whose initial condition, unlike the Glauber one used in section~\ref{sec:matching}, depends on the number of participants. Finally, we evaluate the effectiveness of our matching procedure using a Hard Thermal Loop \cite{Aurenche:2002pd} instead of a Yukawa parton-medium interaction model in appendix~\ref{sec:appHTL}.

\section{Fully-resummed spectrum in evolving media}
\label{sec:full}

We start our discussion with the medium-induced $\vec{k}$-differential spectrum off a high-energy parton, with color representation $C_{\rm R}$ in the BDMPS-Z framework \cite{Baier:1996kr, Baier:1996sk, Zakharov:1996fv,Zakharov:1997uu}. Assuming the medium can be characterized by a time-dependent linear density of scattering centers $n(s)$ with $s\, \in \,[0,L]$, the spectrum in the soft limit for the emitted gluon can be written as \cite{Andres:2020vxs}\footnote{We will use throughout this manuscript bold font for 2D vectors in the transverse plane (with respect to the propagation of the emitter). We adopt the shorthand $\int_{\vec{p}}= \int \mathrm{d}^2\vec{p}/(2\pi)^2$ for the transverse integrals in momentum space.}
\beq
    \omega \frac{\mathrm{d}I^{\mathrm{med}}}
    {\mathrm{d} \omega \mathrm{d}^2 \vec k}=
    \frac{ \alpha_{\rm s}C_{\rm R}}{2 \pi^2 \omega} 
    \operatorname{Re} 
    \int_0^L \mathrm{d}s \, n(s) 
    \int_0^s \mathrm{d}t 
    \int_{\vec p \vec q \vec l} 
    i \vec{p} \cdot \left(
    \frac{\vec l}{\vec l^2}-\frac{\vec q}{\vec q^2}
    \right) 
    \sigma(s;\vec l-\vec q ) 
    \widetilde{\cK}(s, \vec q ; t, \vec p) 
    \cP(L, \vec k ; s, \vec l)\,,
\label{eq:full_kspec}
\eeq
where the vacuum contribution has already been subtracted, and $\omega$ and $\vec{k}$ are, respectively, the radiated gluon energy (assumed to be much smaller than that of the emitter) and its transverse momentum. The dipole cross section $\sigma$ encodes the specific details of the parton-medium interaction model through the elastic collision rate $V(s;\vec{q})$
\beq
\sigma(s;\vec{k}) = - V(s;\vec k) + (2\pi)^2\delta^{(2)}(\vec k)
\int_{\vec q}V(s;\vec q)\, .
\label{eq:sigmaV}
\eeq
Throughout this manuscript, we will use the collision rate for a Yukawa-type elastic parton-medium scattering (also known as Gyulassy-Wang model \cite{Gyulassy:1993hr}) given by 
\beq
V( s;\vec q) = \frac{8 \pi \mu^2(s)}{\left( \vec q^2 + \mu^2(s) \right)^2}\,,
\label{eq:Yukawa}
\eeq
where $\mu^2(s)$ is the time-dependent screening mass of the thermal medium. To illustrate the flexibility of our approach, we present results for the collision rate derived from Hard Thermal Loop (HTL) calculations \cite{Aurenche:2002pd} in appendix~\ref{sec:appHTL}.

In \eqref{eq:full_kspec}, $\cP(L,\vec k;s,\vec l)$ and $\widetilde\cK (s,\vec q;t,\vec p )$ denote, respectively, the transverse momentum broadening and emission kernel in momentum space, which satisfy the following differential equations~\cite{Andres:2020vxs}
\beq
\partial_t \cP(t,\vec k;s,\vec l) = 
-\frac{n(t)}{2}\int_{\vec k'}\, \sigma(t;\vec k-\vec k')\cP(t,\vec k ';s,\vec l)\,,
\label{eq:diffP}
\eeq
\beq
\partial_ t\widetilde{\cK}(s,\vec q;t,\vec p) = \frac{i\vec p^2}{2\omega}
\widetilde{\cK}(s,\vec q;t,\vec p )
+\frac{n(t)}{2}\int_{\vec k'}\sigma(t;\vec k'-\vec p)
\widetilde{\cK}(s,\vec q;t,\vec k')\,,
\label{eq:diffK}
\eeq
with initial conditions given by 
\beq
\cP(s,\vec k;s,\vec l) = (2\pi)^2 \delta^{(2)}(\vec k - \vec l)\,,
\label{eq:ICP}
\eeq
\beq
\widetilde\cK (s,\vec q;s,\vec p ) = (2\pi)^2 \delta^{(2)}(\vec q - \vec p)\,.
\label{eq:ICK}
\eeq

Integrating over the transverse momentum in~\eqref{eq:full_kspec}, one obtains the energy distribution\footnote{We denote by $k=|\vec k|$ the magnitude of 2D vectors in the transverse plane. }
\beq
\omega \frac{\mathrm{d}I^\mathrm{med}}{\mathrm{d}\omega} =
\int_0^\omega k \,\mathrm{d} k 
\int_0^{2\pi} \mathrm{d} \theta_k\,
\omega \frac{\mathrm{d}I^{\mathrm{med}}}{\mathrm{d} \omega \mathrm{d}^2 \vec k}\,,
\label{eq:fullspec_rfin}
\eeq
where we have implemented the kinematic condition restricting the transverse momentum of the emitted gluon to be smaller than its energy $ k \leq \omega$. Although the derivation of the BDMPS-Z spectrum assumes that the transverse momentum of the radiated gluon is significantly smaller than its energy, we also explore the scenario where this kinematic constraint is lifted by extending the integration over $\vec k$ to encompass the entire transverse momentum phase space ($0 \leq k < \infty$). In this case, the broadening factor is integrated out and no longer plays a role in the evaluation of the spectrum. This leads us to the following expression
\beq
    \omega \frac{\mathrm{d}I^{\mathrm{med}}}{\mathrm{d} \omega }=
    \frac{2 \alpha_{\rm s} C_{\rm R}}{\omega} 
    \operatorname{Re} 
    \int_0^L \mathrm{d}s \, n(s) 
    \int_0^s \mathrm{d}t 
    \int_{\vec p \vec q \vec l} 
    i \vec{p} \cdot\left(\frac{\vec l}{\vec l^2}-
    \frac{\vec q}{\vec q^2}\right)  \sigma(s;\vec l-\vec q ) \,
    \widetilde{\cK}(s, \vec q ; t, \vec p) \,.
\label{eq:fullspec_Rinf0}
\eeq
Since $\widetilde \cK$ does not depend on $l$, the integral over $l$ of the second term in brackets in \eqref{eq:fullspec_Rinf0} is zero and we arrive at
\beq
    \omega \frac{\mathrm{d}I^{\mathrm{med}}}{\mathrm{d} \omega }=
    \frac{2 \alpha_{\rm s} C_{\rm R}}{\omega}
    \operatorname{Re} 
    \int_0^L \mathrm{d}s \, n(s) 
    \int_0^s \mathrm{d}t 
    \int_{\vec p \vec q \vec l} 
     i \,\frac{ \vec{p} \cdot\vec l}{\vec l^2}\,
    \sigma(s;\vec l-\vec q ) \,
    \widetilde{\cK}(s, \vec q ; t, \vec p) \,.
\label{eq:fullspec_Rinf}
\eeq

The method for evaluating these emission spectra, \eqref{eq:full_kspec}, \eqref{eq:fullspec_rfin}~and~\eqref{eq:fullspec_Rinf}, was presented in detail in ref.~\cite{Andres:2020vxs}. This involved solving numerically the differential equations that define the in-medium propagators \eqref{eq:diffP}~and~\eqref{eq:diffK} for realistic collision rates, such as the Yukawa \cite{Gyulassy:1993hr} and HTL \cite{Aurenche:2002pd} parton-medium interaction models. However, the numerical evaluation conducted in \cite{Andres:2020vxs,Andres:2020kfg} was limited to static media, assuming the ``brick'' configuration with a longitudinal extension $L$, constant linear density $n(s)=n_0\,\Theta(L-s)$, and constant screening mass $\mu^2(s)=\mu_0^2\,\Theta(L-s)$. As such, the static emission spectrum only depends on the following three medium parameters: $n_0$, $L$, and $\mu_0^2$. In this manuscript, for convenience, we instead adopt the following 
\beq
\chi \equiv n_0L 
\,, \quad
\bar{\omega}_c \equiv \frac{1}{2}\mu_0^2L
 \,, \quad {\rm and}\quad
\bar R \equiv \bar{\omega}_c L\,,
\label{eq:variables_static}
\eeq
where the latter can be regarded as a dimensionless kinematic constraint on the transverse momentum of the radiated gluon, imposing that $k \leq \omega$. Notably, taking the limit $\bar R \rightarrow \infty$ with fixed $\bar \omega_c$ is equivalent to allowing the integration over the transverse momentum of the emitted gluon to extend up to infinity.

Beyond the static case, evaluating the in-medium radiation spectrum for a specific medium profile remains feasible within this approach. However, when considering all possible trajectories of hard partons propagating through the medium, each with a distinct evolution of $n(s)$ and $\mu^2(s)$, the computational time required to solve the involved differential equations (eqs.~\eqref{eq:diffP} and \eqref{eq:diffK}) increases significantly, making it impractical for phenomenological analyses. This is a common challenge inherent to all formalisms aiming to resum all multiple scatterings \cite{Salgado:2002cd}. 

To address this challenge, a well-known workaround, initially proposed in \cite{Salgado:2002cd} for soft emissions within the harmonic oscillator approximation, is to establish a correspondence between the gluon radiation spectrum in a realistic expanding medium and an equivalent static scenario. This matching relation, known as a \emph{scaling law}, enables the utilization of pre-computed static spectra for time-dependent trajectories within a hot QCD medium, thereby substantially reducing the required computational time. More recent works studied refined versions of this scaling to account for emissions beyond the soft-limit \cite{Adhya:2019qse,Adhya:2021kws}. However, it should be noted that all these approaches are limited to the resummation of multiple scatterings within the HO approximation, which fails to reproduce the correct behavior of the large-$\omega$ tails of the in-medium spectrum \cite{Andres:2020vxs}. 

In the following, we will instead focus on the fully resummed spectrum beyond the harmonic oscillator approximation, as derived in \cite{Andres:2020vxs}, analysing different possible matching schemes for this all-order spectrum within a realistic Yukawa parton-interaction model.\footnote{We note that a Monte Carlo attempt of mimicking the all-order in medium rates in \cite{Caron-Huot:2010qjx} was performed in \cite{Park:2016jap,Park:2021yck}.} We also note that we will always consider  the emitter to be a quark, thus taking $C_{\rm R} = C_{\rm F} = 4/3$, and we fix the strong coupling  to $\alpha_{\rm s}=0.3$.

\section{Matching the spectrum for arbitrary trajectories}
\label{sec:matching}

In order to compute the all-order spectrum off a hard parton along its path $\xi(t)$ within a longitudinal evolving medium, one needs as input the relation between the linear density of scattering centers $n(t)$ and screening mass $\mu^2(t)$, and the local properties of the medium. At this point, it is worth remembering that the use in this formalism of the linear density $n$ is motivated by the assumption of homogeneity in the transverse plane,\footnote{Attempts to go beyond this assumption can be found in \cite{Sadofyev:2021ohn,Barata:2022krd,Barata:2023qds} but are not yet available for numerical studies.} in which case the volume density $\rho$ can be replaced by the linear density $n(t)=\rho(t)\sigma_{el}$, with $\sigma_{el}$ the total elastic cross-section which is proportional to the inverse screening mass squared. With this in mind, we establish, along the trajectory of the radiating parton parametrized by $\xi(t)$, the relation between the medium parameters and the linear density and screening mass to be
\beq
n_{\mathrm{hydro}}(t) = k_1 T_{\rm hydro}(\xi(t))\,,  
\qquad  {\rm and } \qquad
\mu_{\mathrm{hydro}}^2(t) = k_2 T^2_{\rm hydro}(\xi(t))\,, 
\label{eq:hydro}
\eeq
where the temperature along the emitter's trajectory $T_{\rm hydro}$ is usually taken from a hydrodynamical simulation of the medium. The proportionality of the screening mass to the temperature in \eqref{eq:hydro} is motivated by the leading order result in the hard thermal loop perturbation theory \cite{Aurenche:2002pd}, while that of the linear density is based on the Stefan-Boltzmann limit of the QCD equation-of-state, where the volume density is proportional to $T^3$. In phenomenological studies, the parameters $k_1$ and $k_2$ would be determined by fitting the experimental data for a specific centrality class. Therefore, for consistency, we will assume in the following that the values of $k_1$ and $k_2$ are the same for all trajectories sampled over a given centrality class. 

In this section, the local temperature $T_{\rm hydro}(\xi(t))$ is extracted from the smooth-averaged 2+1 viscous hydrodynamic model developed in~\cite{Luzum:2008cw,Luzum:2009sb}. This simulation employs as initial condition an energy density proportional to the density of binary collisions, while the ratio of shear viscosity to entropy density is fixed to a constant value of $\eta/s$=0.08. The simulation begins at an initial proper time of $\tau_{\rm hydro}=1\,\rm{fm/c}$ and employs an equation of state inspired by Lattice QCD calculations. The system is assumed to be in chemical equilibrium until it reaches a freeze-out temperature of $T_{\rm f} = 140$ MeV. To compute the radiation spectrum, we will consider different representative straight-line trajectories for various centrality classes at $\sqrt{s_{\rm NN}}=5.02~\rm{TeV}$ Pb-Pb collisions \cite{ALICE:2013hur,ALICE:2015juo}. 

Refs.~\cite{Luzum:2008cw,Luzum:2009sb} further provide a hydrodynamic simulation with the initial condition derived in the Kharzeev-Levin-Nardi (KLN) $k_T$-factorization Color-Glass-Condensate approach~\cite{Drescher:2006pi}. Unlike the Glauber model, the initial condition on the KLN hydrodynamic simulation depends also on the number of participants. We have verified that when using this alternative simulation, our outcomes remain unchanged. Results on the emission spectrum along straight-line trajectories sampled from the KLN simulation are provided in appendix~\ref{sec:appKLN}. 

Finally, we we refer the reader to section~\ref{sec:ebye} for results on the emission spectrum obtained using straight-line trajectories retrieved from the EKRT event-by-event (EbyE) hydrodynamic simulations in \cite{Niemi:2015qia}.

\subsection{Comparison to the static case using average values}
\label{subsec:averages}

\begin{figure}
\includegraphics[width=\textwidth]{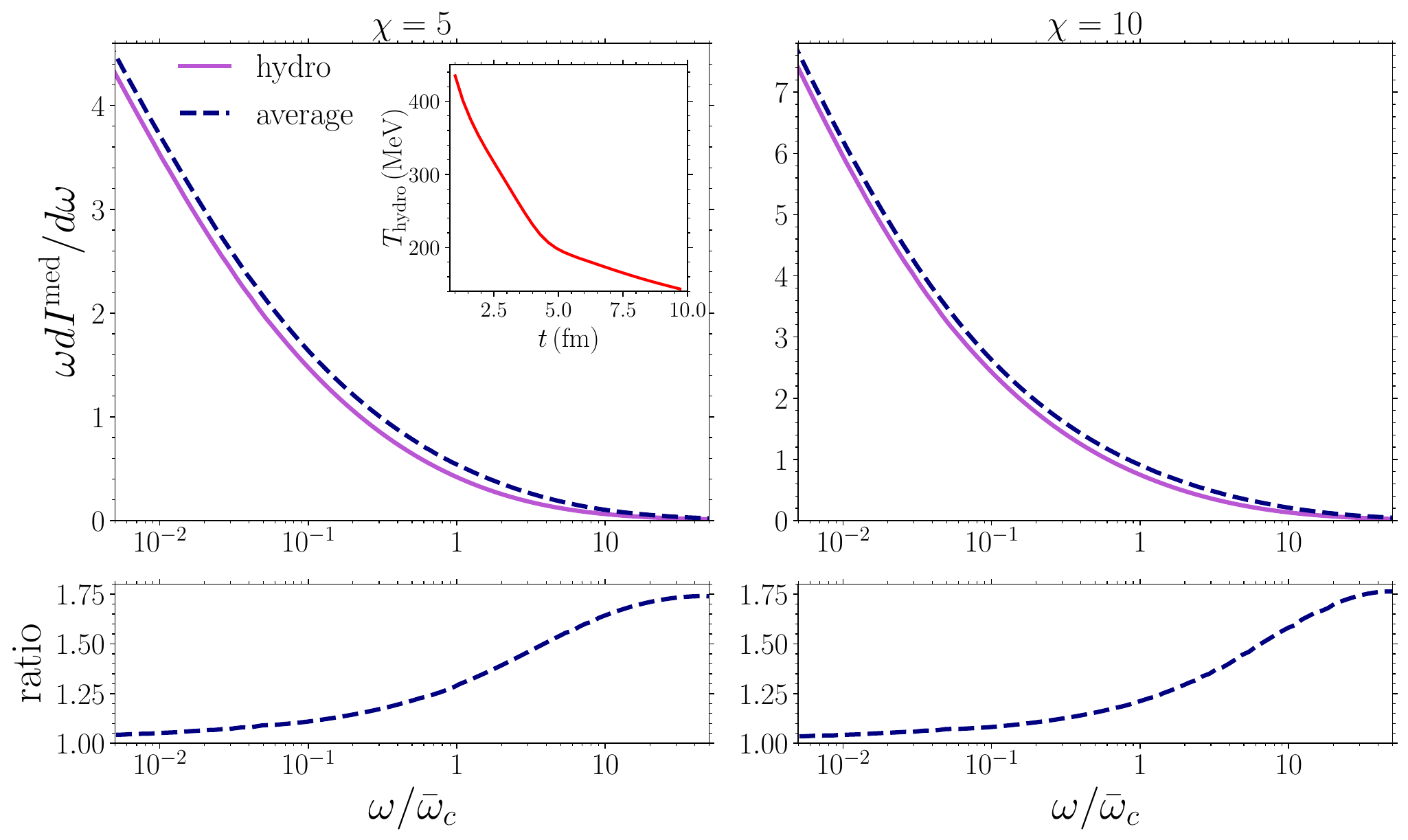}
\caption{Top: fully resummed medium-induced energy distribution for a Yukawa-type interaction with $\bar R \rightarrow \infty$, and $\chi=5$ (left panel) or $\chi=10$ (right panel) as a function of $\omega/\bar{\omega}_c$. The purple curves correspond to the spectra along the trajectory shown in the inset figure, which was sampled with a central production point over the 0-10$\%$ centrality class in $\sqrt{s_{\rm NN}} =5.02$ TeV Pb-Pb collisions at the LHC. For detailed information about the employed path, see main text. The blue dashed curves refer to the static results obtained using average values for the medium parameters (see eqs.~\eqref{eq:average_chi}~and~\eqref{eq:average_omega_c}). Bottom: ratio of the static spectrum w.r.t. the spectrum along the path.}
\label{fig:LHC2_b3.2_x0y0theta0_average}
\end{figure}

Given the availability of results for the emission spectrum in the static case, the first natural step is to seek suitable values for the parameters for the static evaluation that can approximate the behavior of the spectrum in an expanding medium. The first obvious attempt to make (used in several phenomenological studies) is to take the average values of the medium parameters along the emitter's trajectory. For simplicity, let us consider the case without any kinematical constraint \eqref{eq:fullspec_Rinf}, which corresponds to $\bar R\to\infty$ for the static parameters. By taking average values, the resulting static parameters (see eq.~\eqref{eq:variables_static}) are given by:
\beq
\chi = \int_{\tau_{\rm hydro}}^{L+\tau_{\rm hydro}} \mathrm{d}t \,n_{\mathrm{hydro}}(t)\,,
\label{eq:average_chi}
\eeq
\beq
\bar\omega_c = 
\frac{1}{2}
\int_{\tau_{\rm hydro}}^{L+\tau_{\rm hydro}} \mathrm{d}t \,\mu^2_{\mathrm{hydro}}(t)\,,
\label{eq:average_omega_c}
\eeq
where $L$ is the length of the trajectory and $\tau_{\text{hydro}}$ is the initial proper time of the hydrodynamic simulation. These two parameters are the only input needed to evaluate the static energy spectrum given in eq.~\eqref{eq:fullspec_Rinf}.

Figure~\ref{fig:LHC2_b3.2_x0y0theta0_average} illustrates a comparison between the spectrum calculated for a representative trajectory with a central production point sampled over the 0-10$\%$ centrality class in $\sqrt{s_{\rm NN}} =5.02$ TeV Pb-Pb collisions at the LHC (solid purple curve) and the equivalent static scenario (blue dashed curve) defined using the eqs.~\eqref{eq:average_chi}~and~\eqref{eq:average_omega_c}. The spectra are plotted as a function of $\omega/\bar \omega_c$, where $\bar \omega_c$ is determined through eq.~\eqref{eq:average_omega_c}. 
Specifically, the trajectory corresponds to a hard parton propagating from its production point at $\tau_{\rm hydro}$, which is at the midpoint between the centers of the two lead nuclei, along a straight-line along in the in-plane direction. The inset figure showcases the temperature variation over time along this sampled trajectory. We present results for two values of the parameter $k_1$: $k_1=0.5$ (left panel) and $k_1=1.0$  (right panel), which correspond, for the selected trajectory, according to eq.~\eqref{eq:average_chi}, to the static parameters $\chi=5$ and $\chi=10$, respectively. It is evident that the static spectrum, obtained by using average values, overestimates the distribution computed along the trajectory, particularly for larger values of the gluon energy $\omega$. Although we have focused on the case without any kinematical constraint ($\bar R\to \infty$), it is apparent that these significant differences will persist when the kinematical constraint is imposed, since the high-energy tail of the spectrum remains unmodified under the constraint \cite{Salgado:2003gb, Andres:2020vxs}.

We have further analyzed other trajectories sampled with different production points and across different centrality classes in $\sqrt{s_{\rm NN}} =5.02$ TeV Pb-Pb collisions at the LHC, while considering a wide variety of  values of the medium parameters. The corresponding results are very similar to those presented in figure~\ref{fig:LHC2_b3.2_x0y0theta0_average}, which effectively represents the outcomes obtained with the extensive selection of medium parameters. It clearly demonstrates that, when average values are used, the static spectrum does not correctly describe the result obtained along the in-medium path, particularly for large gluon energies. This behavior had already been observed in the context of the harmonic oscillator approximation \cite{Salgado:2003gb}. In the following sections, we will show how the high-energy tails can be appropriately matched.

\subsection{High-energy tail}
\label{subsec:high-energy-tail}
Let us now take a closer look at the emission spectrum in the large-$\omega$ limit for arbitrary functional forms of the linear density $n(t)$ and screening mass $\mu^2(t)$. Firstly, it is important to note that in this kinematic region, the kinematical constraint $k\leq \omega$ is irrelevant, thereby enabling us to safely use eq.~\eqref{eq:fullspec_Rinf}. Secondly, in this high-$\omega$ regime, the first order in opacity contribution dominates the emission spectrum \cite{Mehtar-Tani:2019tvy,Andres:2020vxs}, and thus one can replace the kernel in~\eqref{eq:fullspec_Rinf} by its vacuum version, given by
\beq
\widetilde\cK(s,\vec{q};t,\vec{p})\approx (2\pi)^2\delta^{(2)}(\vec{q}-\vec{p})\,e^{-i\frac{\vec{p}^2}{2\omega}(s-t)}\,,
\eeq
to arrive at
\beq
    \omega \frac{\mathrm{d}I^{\mathrm{med}}}{\mathrm{d} \omega }\approx
    \frac{2 \alpha_{\rm s} C_{\rm R}}{\omega}
    \operatorname{Re} 
    \int_0^L \mathrm{d}s \, n(s) 
    \int_{\vec p \vec l} \frac{2\omega}{\vec{p}^2}\left(1-e^{-i\frac{\vec{p}^2}{2\omega}s}\right) \frac{ \vec{p} \cdot\vec l}{\vec l^2}\,
    \sigma(s;\vec l-\vec p )\,.
\eeq
The integration over $\vec l$ can be easily performed following the appendix of \cite{Andres:2020kfg}, yielding
\beq
\int_{\vec l} \frac{ \vec{p} \cdot\vec l}{\vec l^2}\,\sigma(s;\vec l-\vec p ) = \int_{\vec l}V(s;\vec l)\,\Theta(\vec l^2-\vec p^2) \equiv \Sigma(s;\vec p^2)\,,
\label{eq:Sigma}
\eeq
and thus the emission spectrum is given by
\begin{align}
    \omega \frac{\mathrm{d}I^{\mathrm{med}}}{\mathrm{d} \omega } &\approx
    \frac{2 \alpha_{\rm s} C_{\rm R}}{\omega}
    \operatorname{Re} 
    \int_0^L \mathrm{d}s \, n(s) 
    \int_{\vec p} \frac{2\omega}{\vec{p}^2}\left(1-e^{-i\frac{\vec{p}^2}{2\omega}s}\right)\Sigma(s;\vec p^2)\nonumber\\
    &= \frac{\alpha_{\rm s} C_{\rm R}}{\pi}
    \operatorname{Re} 
    \int_0^L \mathrm{d}s \, n(s) 
    \int_0^\infty \frac{\mathrm{d}r}{r} \left(1-e^{-ir}\right)\Sigma(s;2\omega r/s)\,.
    \label{eq:large_omega0}
\end{align}
For the particular case of the Yukawa parton-interaction model given in eq.~\eqref{eq:Yukawa}, it is possible to compute $\Sigma$ exactly, yielding
\beq
\Sigma(s;\vec p^2) = \frac{2\mu^2(s)}{\vec p^2+\mu^2(s)}\,.
\eeq
Consequently, in the limit of large $\omega$, we obtain 
\beq
\Sigma(s;2\omega r/s)\xrightarrow[\omega \rightarrow \infty]{} \frac{s\,\mu^2(s)}{\omega r}\,,
\eeq
and thus the two integrations in the spectrum \eqref{eq:large_omega0} factorize as
\begin{align}
\omega \frac{\mathrm{d}I^{\mathrm{med}}}{\mathrm{d} \omega } &\approx \frac{\alpha_{\rm s} C_{\rm R}}{\pi\omega}
\operatorname{Re} \int_0^L \mathrm{d}s \,s\, n(s)\, \mu^2(s)
\int_0^\infty \frac{\mathrm{d}r}{r^2} \left(1-e^{-ir}\right)\nonumber\\
&= \frac{\alpha_{\rm s} C_{\rm R}}{2\omega} \int_0^L \mathrm{d}s \,s\, n(s)\, \mu^2(s)\,.
\label{eq:largeomega}
\end{align}
It is evident from this expression that the spectrum at large $\omega$ continues to exhibit the well-known $1/\omega$-behavior of the static result, which can be written as
\beq
\label{eq:spec_static}
\left. \omega \frac{\mathrm{d}I^{\mathrm{med}}}{\mathrm{d} \omega}\right|_{\rm static} \approx \frac{\alpha_{\rm s} C_{\rm R}}{2}\frac{\chi \bar \omega_c}{\omega}\,.
\eeq
Furthermore, the time dependence of the medium parameters enters the spectrum in~\eqref{eq:largeomega} only through a specific integration. This allows us to establish the appropriate scaling law to ensure that the large-$\omega$ behavior of the static spectra matches the result computed along the actual path. Importantly, this result is not limited to the specific derivation presented here, but rather holds for any interaction model $V$ that satisfies $V(s;\vec p)\propto\mu^2(s)/\vec p^4$ in the large momentum limit, such as the HTL collision rate \cite{Aurenche:2002pd}. This is due to the fact that for large argument values, $\Sigma$ depends solely on the large-momentum tail of $V$.

\subsection{Scaling laws with respect to static media}

By comparing eqs.~\eqref{eq:largeomega}~and~\eqref{eq:spec_static}, one can straightforwardly determine which is the correct combination of parameters that allows the matching of the high-energy tails of both spectra. This scaling law is given by:
\beq
\chi\, \bar\omega_c = \int_{\tau_{\rm hydro}}^{L+\tau_{\rm hydro}} \mathrm{d}t\,t\,n_{\rm hydro}(t)\,\mu^2_{\rm hydro}(t)\,,
\label{eq:1moment_static}
\eeq
where we have shifted the integration limits on the right-hand side by $\tau_{\rm hydro}$, since this is the initial time of the hydrodynamics simulations. We note that this relationship had already been conjectured in \cite{Salgado:2003gb} for two particular cases: the harmonic approximation with $\hat q(t)\sim n(t)\mu^2(t)$, and the first order in opacity (GLV) with a constant screening mass.

In order to obtain a complete matching scheme with the static case, it is necessary to establish two additional relations that allow the unambiguous extraction of the three static parameters in \eqref{eq:variables_static}. Following \cite{Salgado:2003gb,Andres:2016iys}, we adopt 
\begin{align}
    \chi &= \int_{\tau_{\rm hydro}}^{L+\tau_{\rm hydro}}\mathrm{d}t\,n_{\rm hydro}(t)\,,\\
    \chi\,\bar R &= \frac{3}{2}\int_{\tau_{\rm hydro}}^{L+\tau_{\rm hydro}}\mathrm{d}t\, t^2\,n_{\rm hydro}(t)\,\mu^2_{\rm hydro}(t)\,.
    \label{eq:2moment_static}
\end{align}
It is worth noting that when considering the case without the kinematical condition, the latter relation becomes irrelevant as $\bar R \rightarrow \infty$.

\begin{figure}
\includegraphics[width=\textwidth]{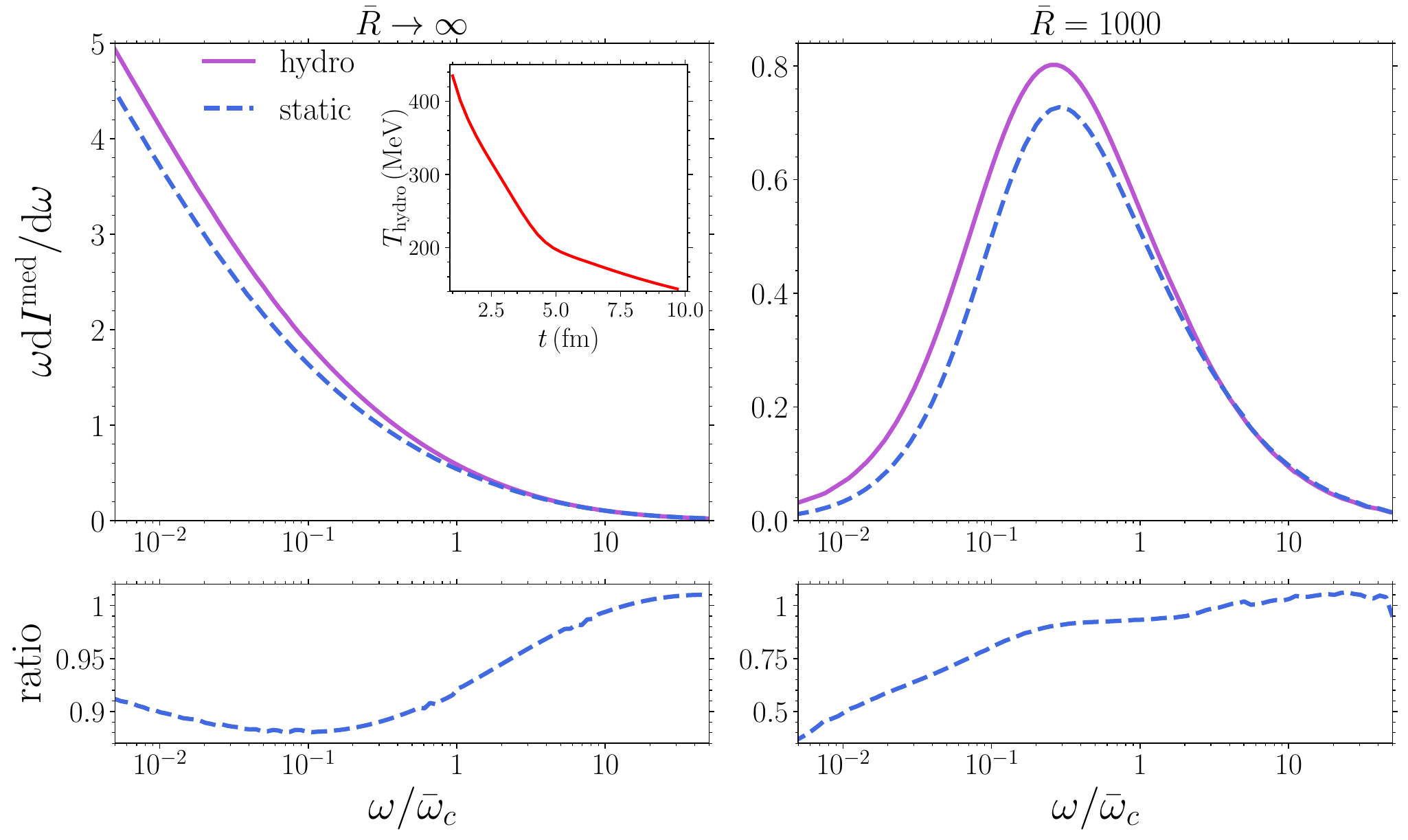}
\caption{Top: fully resummed medium-induced energy distribution for the Yukawa-type interaction for $\chi = 5$, and $ \bar{R}  \rightarrow \infty$ (left panel) or $\bar{R} = 1000$ (right panel) as a function of $\omega/\bar{\omega}_c$. The purple curves correspond to the spectra computed along the temperature profile shown in the inset figure and sampled with a central production point over the 0-10$\%$ centrality class in $\sqrt{s_{\rm NN}} =5.02$ TeV Pb-Pb collisions at the LHC. For detailed information about the employed path, see main text. The blue dashed curves refer to the results for the static scenario given by eqs.~\eqref{eq:1moment_static}-\eqref{eq:2moment_static}. Bottom: ratio  of the static spectrum w.r.t. the spectrum along the path.}
\label{fig:LHC2_b3.2_x0y0theta0_static}
\end{figure}

In figure~\ref{fig:LHC2_b3.2_x0y0theta0_static}, we present the resulting energy distributions for the static case (blue dashed curve) and the hydrodynamically evolving scenario (solid purple curve) as a function of $\omega/\bar{\omega}_c$. These results were matched using the scaling laws \eqref{eq:1moment_static}-\eqref{eq:2moment_static} for the same  trajectory employed in figure~\ref{fig:LHC2_b3.2_x0y0theta0_average} --- hard parton produced at the midpoint between the centers of the two lead nuclei and moving along the in-plane direction. The  temperature profile along this path as a function of time is shown in the inset panel of this figure. In both panels, we present results using $\chi =5$ (or, equivalently, $k_1=0.5$), with the left panel corresponding to $\bar{R} \rightarrow \infty$ limit and the right panel to $\bar R=1000$ (or, equivalently, $k_2=45$). We observe that the proposed scaling laws, in contrast to the average values shown in figure~\ref{fig:LHC2_b3.2_x0y0theta0_average}, yield a static spectrum that accurately reproduces the high-$\omega$ tails of the spectrum computed along the trajectory. However, for lower gluon energies, there are still significant discrepancies between the static and hydrodynamically evolving scenarios.

Overall, the results shown in figure~\ref{fig:LHC2_b3.2_x0y0theta0_static} are representative of the performance of the matching given by eqs.~\eqref{eq:1moment_static}-\eqref{eq:2moment_static}. Results for other trajectories with other medium parameters will be presented in subsequent sections where we will provide a comprehensive analysis of this and another matching scheme.

\subsection{Scaling laws with respect to a power-law profile}
\label{subsec:power}

The main objective of adopting scaling laws is to enable the pre-computation of the in-medium spectrum for various parameter combinations, which can then be applied to a wide range of realistic evolving scenarios. For instance, one could pre-evaluate the static spectrum by considering different variations of its three parameters given in   eq.~\eqref{eq:variables_static}. To employ these pre-tabulated spectra in phenomenological analyses, it would be sufficient to determine the values of the parameters along each trajectory sampled from a hydrodynamic simulation using equations \eqref{eq:1moment_static}-\eqref{eq:2moment_static}.

However, it is important to note that this static matching, as depicted in figure~\ref{fig:LHC2_b3.2_x0y0theta0_static}, yields significant errors. To mitigate these errors, we propose an alternative approach: finding an equivalent scenario characterized by a power-law evolving medium. This power-law scaling allows for the pre-computation of the spectrum for a selected set of parameters, which can accurately describe the spectrum for a wide range of realistic scenarios within a hydrodynamically evolving QGP. By employing this new scaling, we aim at improving the accuracy and applicability of the spectra in phenomenological studies.

In this approach, we express the linear density of scattering centres $n(t)$ and the screening mass $\mu(t)$ as follows
\beq
n(t)=\frac{\bar{n}_0}{(t+\bar{t}_0)^{\alpha}}\,,
\quad 
\mathrm{and}
\quad
\mu^2(t)=\frac{\bar{\mu}_0^2}{(t+\bar{t}_0)^{2\alpha}}\,,
\label{eq:powerlaw}
\eeq
with $\bar{n}_0$ and $\bar{\mu}_0$ proportional to the maximum density and screening mass, respectively, at the initial time $\bar{t}_0$. In the case of $\alpha = 1$, we obtain an evolving medium with free streaming along the longitudinal direction, commonly known as Bjorken expansion. 

The corresponding scaling laws for this power-law scenario are given by
\beq
 \int_0^{\bar{L}} \mathrm{d}t \,n(t)  =
 \int_{\tau_{\rm hydro}}^{L + \tau_{\rm hydro}} \mathrm{d}t \,n_{\mathrm{hydro}}(t) \,,
 \label{eq:0moment_powerlaw}
 \eeq
 \beq
\int_0^{\bar{L}} \mathrm{d}t \, t\,n(t)\,\mu^2(t)  =
 \int_{\tau_{\rm hydro}}^{L + \tau_{\rm hydro}} \mathrm{d}t \, t\,n_{\mathrm{hydro}}(t)\,\mu^2_{\mathrm{hydro}}(t)\,,
\label{eq:1moment_powerlaw}
 \eeq
 \beq
 \int_0^{\bar{L}} \mathrm{d}t \, t^2\,n(t)\,\mu^2(t)  
 = \int_{\tau_{\rm hydro}}^{L + \tau_{\rm hydro}} \mathrm{d}t \,t^2\,n_{\mathrm{hydro}}(t)\,\mu^2_{\mathrm{hydro}}(t) \,.
 \label{eq:2moment_powerlaw}
 \eeq
where $\bar L$ represents the endpoint of the trajectory along the power-law scenario and \eqref{eq:1moment_powerlaw} ensures that the high-$\omega$ tail of the power-law spectrum matches that of the spectrum computed along the actual hydrodynamic path, as described in section~\ref{subsec:high-energy-tail}.

Using these equations, we can determine the parameters of the power-law evolving medium that best approximate the spectrum along a given hydrodynamic profile $T(\xi(t))$. To maintain the same combination of three parameters as in the static case (see eq.~\eqref{eq:variables_static}), we need to fix the values of $\alpha$ and the dimensionless variable $t_0 = \bar{t}_0/\bar{L}$. In the present analysis, we have chosen $\alpha = 0.5$ and $t_0 = 0.1$ as they provided the best approximation to the energy spectrum computed along realistic temperature profiles $T(\xi(t))$. As such, in the following, we will compare this scenario to the static matching approach. For additional results with different values of $\alpha$ and $t_0$, and the $\chi^2$ analysis leading to the choice of $\alpha=0.5$ we refer the reader to appendix \ref{sec:appA}. 

\begin{figure}
\includegraphics[width=\textwidth]{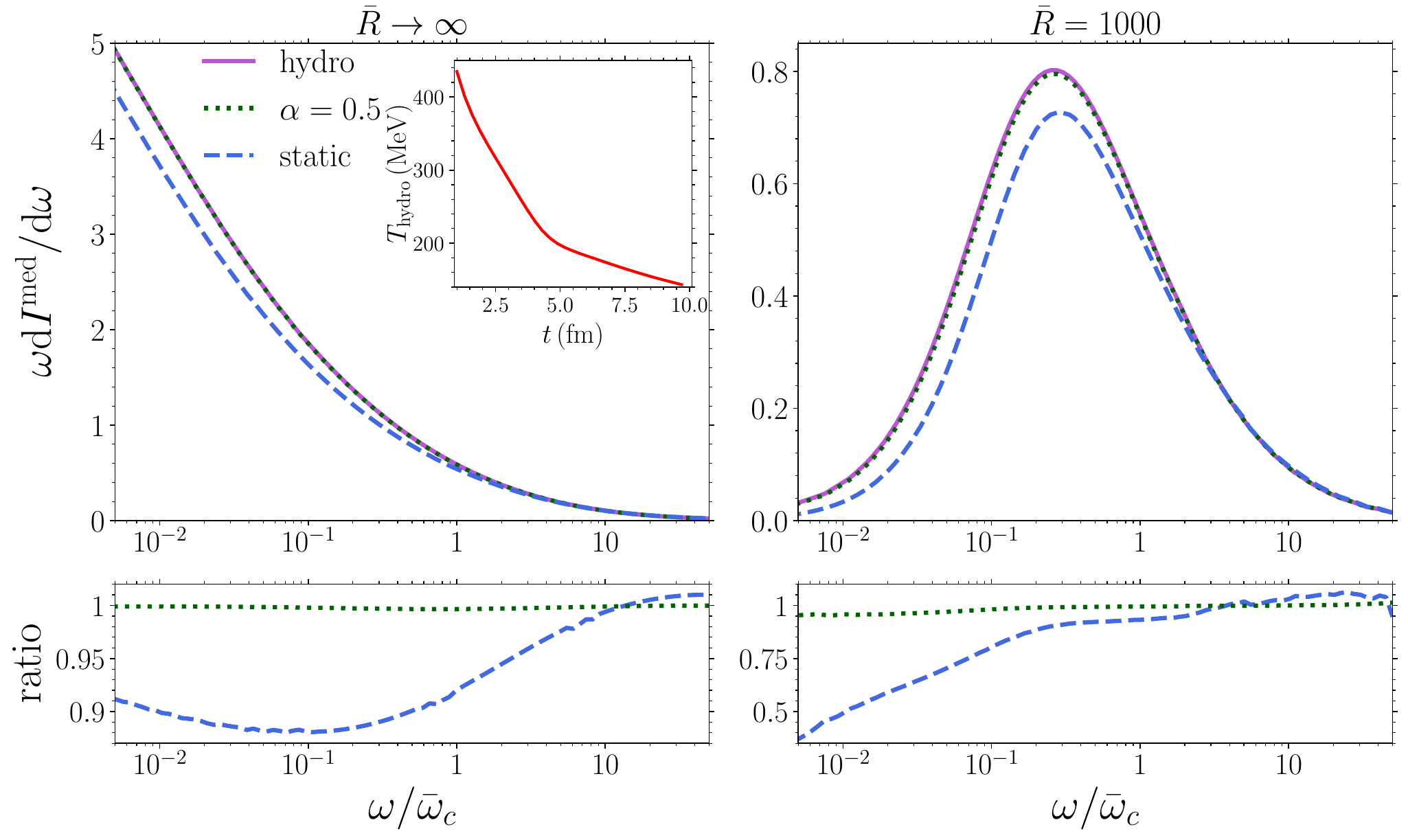}
\caption{Top: fully resummed medium-induced gluon energy distribution for a Yukawa interaction model with $\chi = 5$, and $ \bar R \rightarrow \infty$ (left panel) or $ \bar R  = 1000$ (right panel) as a function of $\omega/\bar \omega_c$. The purple curves correspond to the spectra along the temperature profile shown in the inset figure sampled with a  central production point over the 0-10$\%$ centrality class in $\sqrt{s_{\rm NN}} =5.02$ TeV Pb-Pb collisions at the LHC. For detailed information about the employed path, see main text. The dashed blue and dotted green curves correspond, respectively, to the results for the static (see eqs.~\eqref{eq:1moment_static}-\eqref{eq:2moment_static}) and power-law scenarios (see eqs.~\eqref{eq:0moment_powerlaw}-\eqref{eq:2moment_powerlaw}). Bottom: Ratio of the power-law (static) spectrum w.r.t. the spectrum along the temperature profile in dotted green (dashed blue).}
  \label{fig:LHC2_b3.2_x0y0theta0_a05}
\end{figure}

We present in figure~\ref{fig:LHC2_b3.2_x0y0theta0_a05} the energy distribution computed along the same path (hard parton produced at the midpoint between the centers of the two lead nuclei and moving along the in-plane direction) and with the same parameter values $k_1=0.5$ and $k_2=45$ as in figures~\ref{fig:LHC2_b3.2_x0y0theta0_average}~and~\ref{fig:LHC2_b3.2_x0y0theta0_static} (solid purple curve) as a function of $\omega/\bar \omega_c$, with $\bar \omega_c$ defined through \eqref{eq:1moment_static}. We compare this result to the static scenario obtained through the scaling laws \eqref{eq:1moment_static}-\eqref{eq:2moment_static} (dashed blue curve) and the proposed power-law scaling given by \eqref{eq:0moment_powerlaw}-\eqref{eq:2moment_powerlaw} (dotted green curve). Unlike the static scenario, the power-law scaling yields an accurate description of the spectrum along the hydrodynamic path for all gluon energies, regardless of whether the kinematic constraint is removed (left panel) or imposed (right panel).

\begin{figure}
\includegraphics[width=\textwidth]{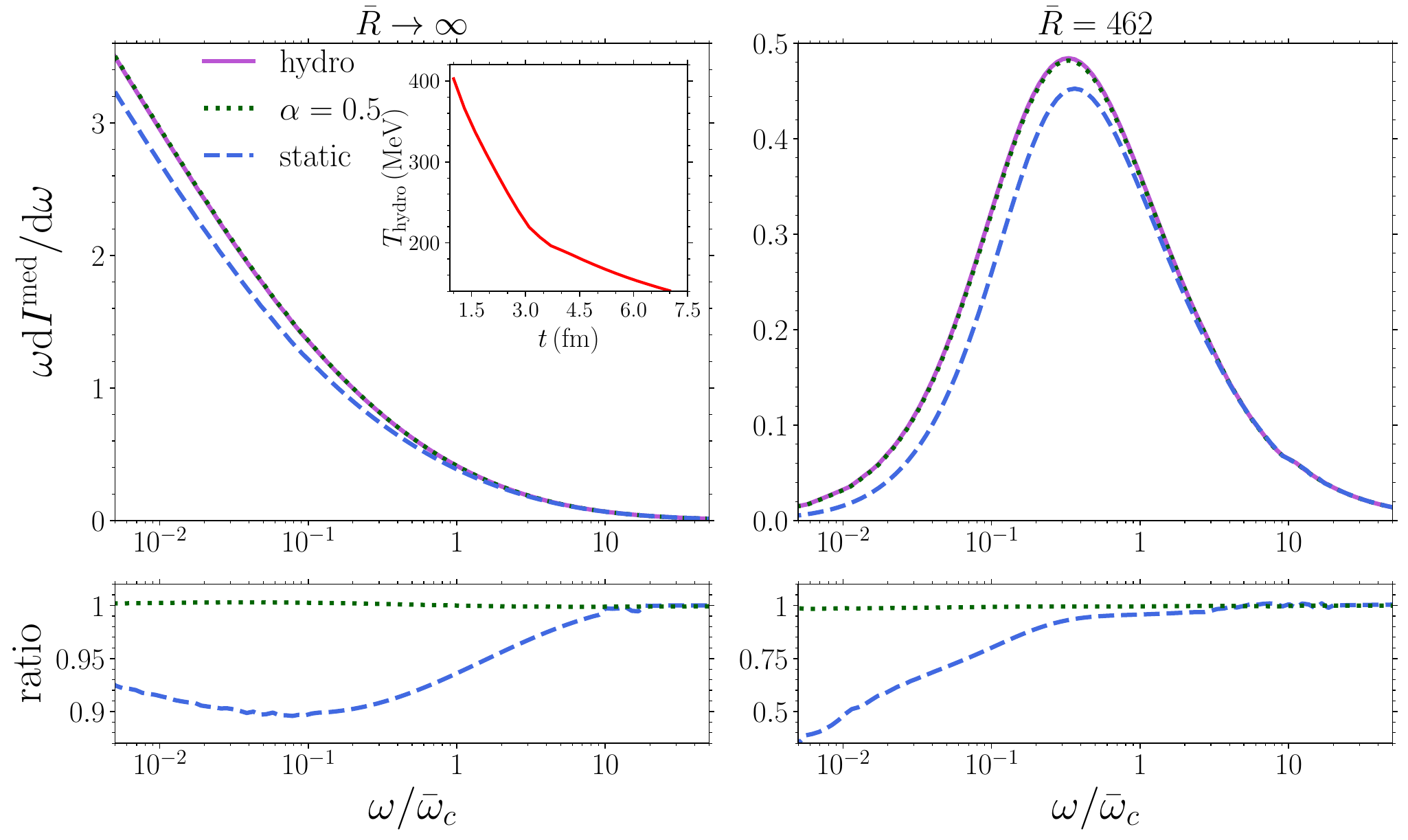}
\caption{Top: fully resummed medium-induced energy distribution for the Yukawa-type interaction with $\chi = 3.3$, and $\bar R \rightarrow \infty$ (left panel) or $ \bar R  = 462$ (right panel) as a function of $\omega/\bar \omega_c$. The purple curves correspond to the spectra along the temperature profile shown in the inset figure sampled over the 0-10$\%$ centrality class in $\sqrt{s_{\rm NN}} =5.02$ TeV Pb-Pb collisions at the LHC with an off-central production point. For detailed information about the employed path, see main text. The dashed blue and dotted green curves correspond, respectively, to the results for the static (see eqs.~\eqref{eq:1moment_static}-\eqref{eq:2moment_static}) and power-law scenarios (see eqs.~\eqref{eq:0moment_powerlaw}-\eqref{eq:2moment_powerlaw}). Bottom: ratio of the power-law (static) spectrum w.r.t. the spectrum along the temperature profile in dotted green (dashed blue).}
\label{fig:LHC2_b3.2_x10y10theta90_a05}
\end{figure}

In figure~\ref{fig:LHC2_b3.2_x10y10theta90_a05}, we present results for a different, but still typical, straight-line trajectory sampled over the 0–10$\%$ centrality class in $\sqrt{s_{\rm NN}} =5.02$ TeV Pb-Pb collisions at the LHC but off-central production point. Specifically, the trajectory of a hard parton produced at $\tau_{\rm hydro}$ 1\,fm away both in the $x$- and $y$-directions from the midpoint between the centers of the two nuclei, and propagating in the out-of-plane direction. The inset figure shows the temperature variation over time along this sampled path. We have used the same parameter values $k_1$ and $k_2$ as in the previous figures, resulting, in terms of the static ones, in $\chi=3.3$ and $\bar R \rightarrow \infty$ (left panel),  and $\chi=3.3$ and $\bar R =462$ (right panel). As in figure~\ref{fig:LHC2_b3.2_x0y0theta0_a05},  the power-law scaling (blue dashed curve)  accurately describes the spectrum along the hydrodynamic path (purple solid curves) for all gluon energies $\omega$.

\begin{figure}
\includegraphics[width=\textwidth]{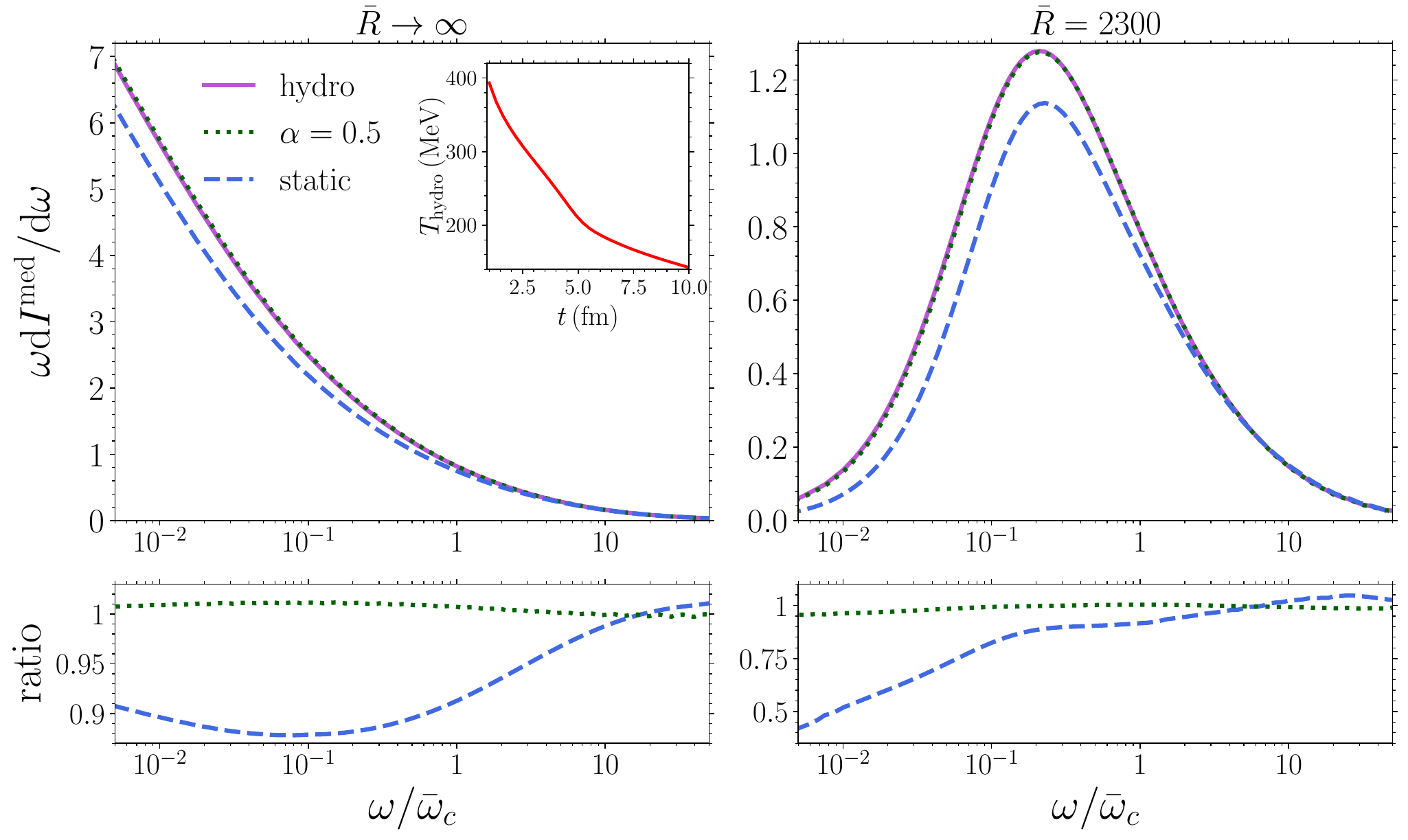}
\caption{Top: fully resummed medium-induced energy distribution for the Yukawa interaction model with  $\chi=7.6$, and $\bar R \rightarrow \infty$ (left panel) or $\bar R= 2300$ (right panel) as a function of $\omega/\bar{\omega}_c$. The purple curves correspond to the spectra along the temperature profile shown in the inset figure sampled over the 20-30$\%$ centrality class in $\sqrt{s_{\rm NN}} =5.02$ TeV Pb-Pb collision at the LHC. For detailed information about the employed path, see main text. The dashed blue and dotted green curves correspond, respectively, to the spectra for the static (see eqs.~\eqref{eq:1moment_static}-\eqref{eq:2moment_static}) and power-law scenarios (see eqs.~\eqref{eq:0moment_powerlaw}-\eqref{eq:2moment_powerlaw}). Bottom: ratio of the power-law (static) spectrum w.r.t. the spectrum along the temperature profile in dotted green (dashed blue).}
\label{fig:LHC2_b7.2_x0y30theta270}
\end{figure}

We have conducted an extensive analysis of numerous straight-line trajectories sampled across several centrality classes in $\sqrt{s_{\rm NN}} = 5.02$ TeV Pb-Pb collisions, considering a wide range of medium parameter values. Our findings consistently demonstrate that for the majority of these trajectories, characterized by temperature profiles that monotonically decrease with time, the power-law scaling approach provides an excellent description of the spectrum along the path across its entire kinematic range. In fact, the ratio between the power-law spectrum and the actual spectrum along the path always remains below $2\%$ for any monotonically decreasing temperature profiles sampled over central to semi-peripheral LHC Pb-Pb collisions. This indicates a remarkable level of agreement between the two spectra, validating the effectiveness of the power-law scaling approach. To further illustrate this point, we present in figure~\ref{fig:LHC2_b7.2_x0y30theta270} the results for a typical path sampled over 20-30$\%$ semi-peripheral $\sqrt{s_{\rm NN}} = 5.02$ TeV Pb-Pb collision at the LHC. Namely, for a the trajectory corresponding to a hard parton produced in the out-of plane direction, 3 fm away from the midpoint between the center of the two nuclei, and propagating with a 270$^{\circ}$ angle with respect to the in-plane direction. The inset panel shows the corresponding temperature profile for this trajectory. In both panels, we present results using $\chi=7.6$ (or, equivalently, $k_1=0.75$), with the left panel corresponding to $\bar R \rightarrow \infty$ and the right panel to $\bar R= 2300$ (or, equivalently, $k_2=93$). We observe that for large gluon energies, both the power-law and static scaling laws accurately reproduce the spectrum along the path, as expected. However, for lower gluon energies, we find that the power-law spectrum and the actual spectrum along the path exhibit excellent agreement, with differences of less than $1\%$, regardless of whether the kinematic cutoff is imposed (right panel) or removed (left panel). In contrast, the static scaling law result deviates significantly from the observed spectra in this energy range. 

\begin{figure}
\includegraphics[width=\textwidth]{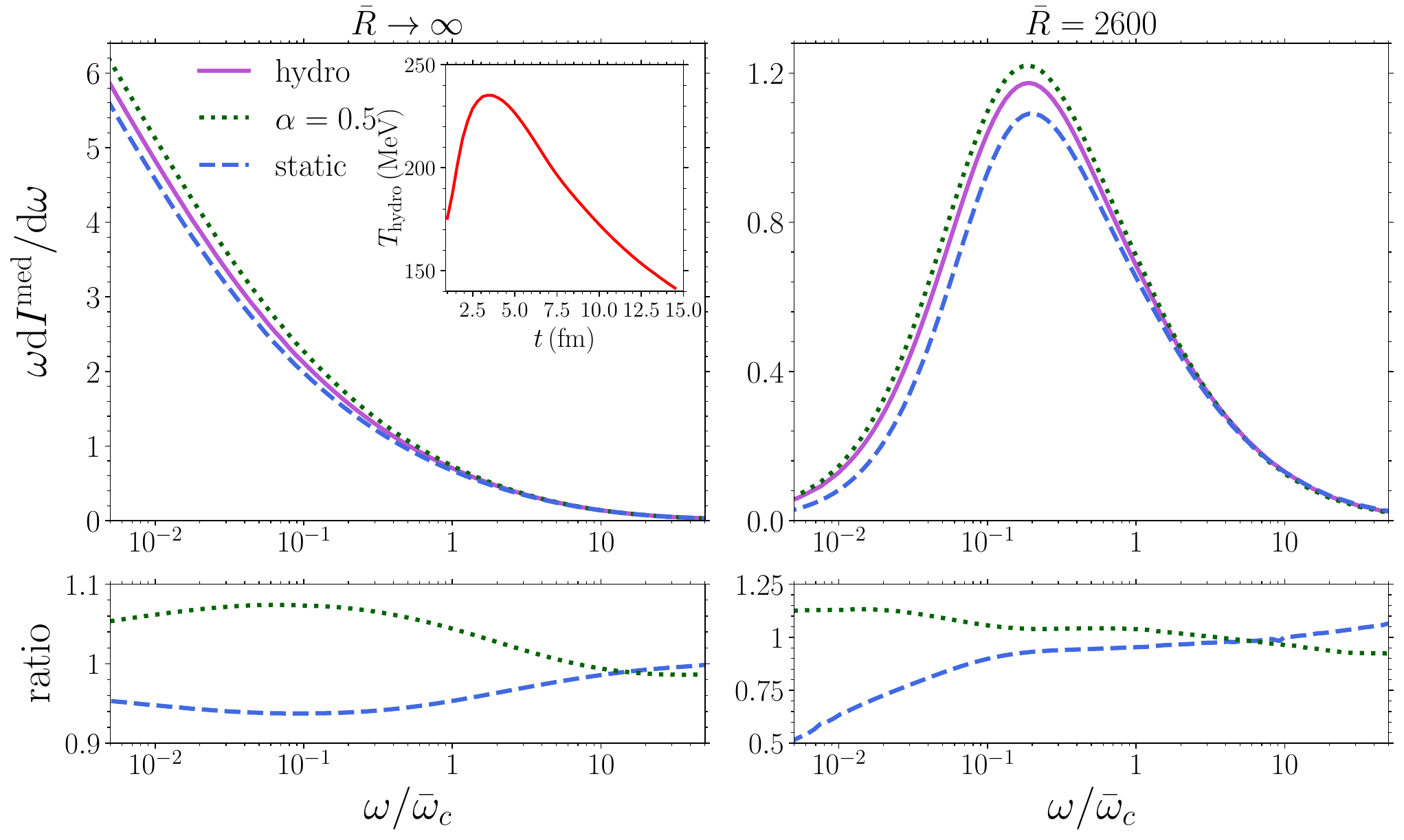}
\caption{ Top: fully resummed medium-induced energy distribution for the Yukawa interaction model with  $\chi=6.6$, and $ \bar R  \rightarrow \infty$ (left panel) or $ \bar R  = 2600$ (right panel) as a function of $\omega/\bar{\omega}_c$. The purple curves correspond to the spectra along the rare temperature profile shown in the inset figure sampled with a non-central production point over the 0-10$\%$ centrality class in $\sqrt{s_{\rm NN}} =5.02$ TeV Pb-Pb collisions at the LHC. For detailed information about the employed path, see main text. The dashed blue and dotted green curves correspond, respectively, to the spectra for the static (see eqs.~\eqref{eq:1moment_static}-\eqref{eq:2moment_static}) and power-law matchings (see eqs.~\eqref{eq:0moment_powerlaw}-\eqref{eq:2moment_powerlaw}). Bottom: ratio of the power-law (static) spectrum w.r.t. the spectrum along the temperature profile in dotted green (dashed blue).}
\label{fig:LHC2_b3.2_x0y80theta255}
\end{figure}

While figures \ref{fig:LHC2_b3.2_x0y0theta0_a05},~\ref{fig:LHC2_b3.2_x10y10theta90_a05}~and~\ref{fig:LHC2_b7.2_x0y30theta270} provide a good representation of the majority of paths that can be sampled from Pb-Pb collisions at LHC energies, it is worth noting that in  rare cases, the hard parton may be produced near the edge of the nuclei overlapping region and propagate a long distance through the medium, resulting in temperature profiles that do not monotonically decrease with time. We have extensively investigated such extreme scenarios, which constitute about $10\%$ of the possible events in the 20-30$\%$ centrality class and about $7\%$ in the 0-10$\%$ centrality class and have found that even in these cases, the power-law scaling law provides a reasonably good approximation of the spectrum along the path, with deviations never exceeding $15\%$. This is illustrated in figure~\ref{fig:LHC2_b3.2_x0y80theta255} for a path sampled over the 0-10$\%$ centrality class in $\sqrt{s_{\rm NN}} =5.02$ TeV Pb-Pb collisions at the LHC, where the temperature profile exhibits a non-monotonic decreasing behavior (as shown in the inset panel). This specific trajectory corresponds to a hard parton produced at $\tau_{\rm hydro}$ 8\,fm away in the out-of-plane direction from the midpoint between the centers of the nuclei, and propagating inwards with a 255$^\circ$ angle with respect to the in-plane direction. We have used in this figure the same parameter values $k_1=0.5$ and $k_2=45$ as in figures~\ref{fig:LHC2_b3.2_x0y0theta0_a05}~and~\ref{fig:LHC2_b3.2_x10y10theta90_a05}, resulting for this path in  $\chi=6.6$ and $\bar R \rightarrow \infty$ (left panel),  and $\chi=6.6$ and $\bar R =2600$ (right panel). We observe that the power-law result starts to deviate from the spectrum along the path for gluon energies smaller than the characteristic gluon energy $\bar \omega_c$. However, it is important to note that these deviations remain below $15\%$ for the entire range of gluon energies. In fact, in this case, the deviations from the power-law scaling approach are not larger than the deviations observed in the static matching scenario. Nonetheless, we would like to emphasize that figure~\ref{fig:LHC2_b3.2_x0y80theta255} depicts the most extreme scenario we have encountered in terms of temperature increase at initial times. Therefore, for any other trajectory, including those sampled from different centralities, the deviations of the power-law result from the spectrum along the path are expected to be smaller. 

Finally, in the calculation of any observable, one needs to account for all possible trajectories, each weighted by the probability of producing the hard parton at its initial point, which can be obtained from the nuclear overlap distribution of the incoming nuclei. Estimating the error of performing an average over a large number of trajectories using power-law scaling spectra instead of actual spectra along the hydrodynamics paths becomes unfeasible due the large computational cost associated with calculating the spectrum along all paths at the high level of accuracy presented here.\footnote{We reemphasize that the extremely high computational cost of performing an average over a large number of trajectories of the hard parton is inherent to pQCD approaches accounting for all-order resummation of multiple scatterings \cite{Zakharov:2004vm,Caron-Huot:2010qjx,Moore:2021jwe,Schlichting:2021idr,Yazdi:2022bru,Feal:2018sml,Andres:2020vxs,Mehtar-Tani:2019tvy,Barata:2020sav,Andres:2020kfg,Andres:2022bql}. In fact, many semi-analytical and Monte Carlo approaches to energy loss have performed such averaging using additional approximations of the emission spectrum or energy loss, which speed up its evaluation (see, for instance, refs.~\cite{Marquet:2009eq,Armesto:2009zi,Renk:2011gj,Wang:2013cia,Andres:2016iys,Cao:2017hhk,Hulcher:2017cpt,Casalderrey-Solana:2018wrw,He:2018xjv,Huss:2020dwe,Huss:2020whe,Zigic:2021rku,Mehtar-Tani:2021fud,JETSCAPE:2021ehl,Xie:2022ght,JETSCAPE:2022jer,Xie:2022fak,Luo:2023nsi,Mehtar-Tani:2024jtd}).} Nevertheless, it is important to note that the most problematic trajectories, the ones for which the scaling is less accurate, would have a very small weight in such an average. Our analysis shows that most trajectories have an error smaller than 2$\%$ when the spectrum is computed using the power-law scaling, while only rare trajectories produced in the periphery of the collision with non-monotonically decreasing temperatures can have an error of up to 15$\%$. Such rare trajectories are produced in regions where the nuclear overlap distribution is very small. For example, the trajectory shown in figure~\ref{fig:LHC2_b3.2_x0y80theta255} has  a relative weight of 0.00006  with respect to the trajectory starting at the mid-point between the nuclei for the same centrality class shown in figure~\ref{fig:LHC2_b3.2_x0y0theta0_a05}. This is contrast with the vast majority of trajectories contributing to the average that behave as the one in figure~\ref{fig:LHC2_b3.2_x10y10theta90_a05}, which has a relative weight of 0.95 with respect to that in figure~\ref{fig:LHC2_b3.2_x0y0theta0_a05} and for which the scaling law gives a maximum error of 2$\%$.

\section{Impact of event-by-event fluctuations}
\label{sec:ebye}

In this section, we analyze the influence of event-by-event fluctuations on the performance of the power-law matching relations described in section~\ref{subsec:power}. For this purpose, we make use of the EbyE EKRT hydrodynamics simulation \cite{Niemi:2015qia}, which includes fluctuating initial energy density profiles obtained within the EKRT framework \cite{Eskola:1999fc} and has an initialization time of $\tau_{\rm hydro} = 0.197$ fm. This simulation employs as equation of state the s95p parametrization of the lattice QCD results \cite{Huovinen:2009yb}, and the shear viscosity is parameterized as $\eta/s(T) = param1$, as described in ref.~\cite{Niemi:2015qia}. It also implements a chemical freeze-out at $T_{\rm chem}=175$ MeV and a kinetic freeze-out at $T_{\rm dec}= 100$ MeV. It is worth noting that the results from this hydrodynamics exhibit excellent agreement with measurements of soft hadronic observables, including multiplicity, average transverse momentum, flow coefficients and flow correlations, in $\sqrt{s_{\rm NN}} = 200~\rm{GeV}$ Au-Au collisions at RHIC, and $\sqrt{s_{\rm NN}} = 2.76~\rm{TeV}$  Pb-Pb collisions at the LHC \cite{Eskola:1999fc}.

\begin{figure}
\includegraphics[width=0.5\textwidth]{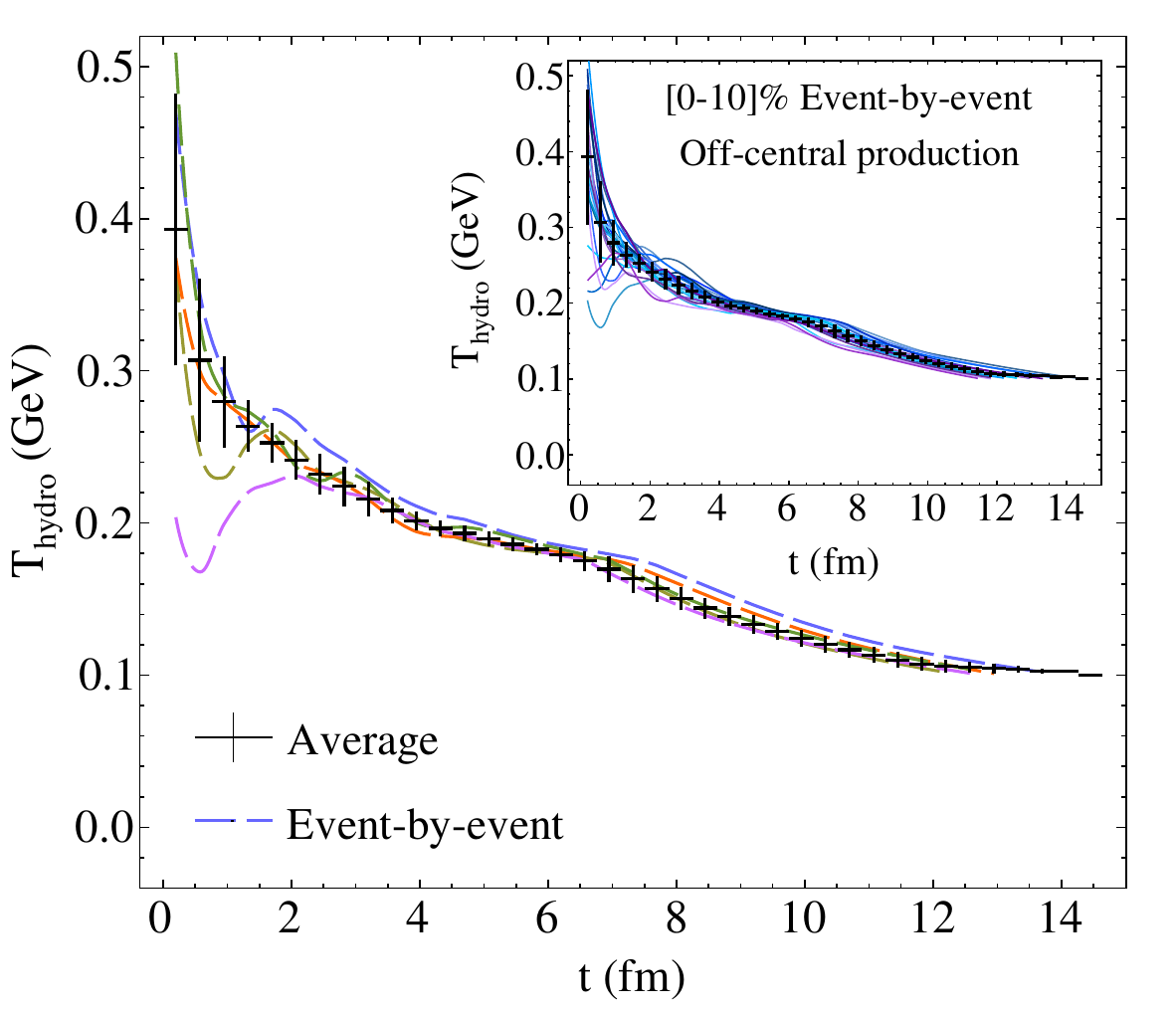}
\includegraphics[width=0.5\textwidth]{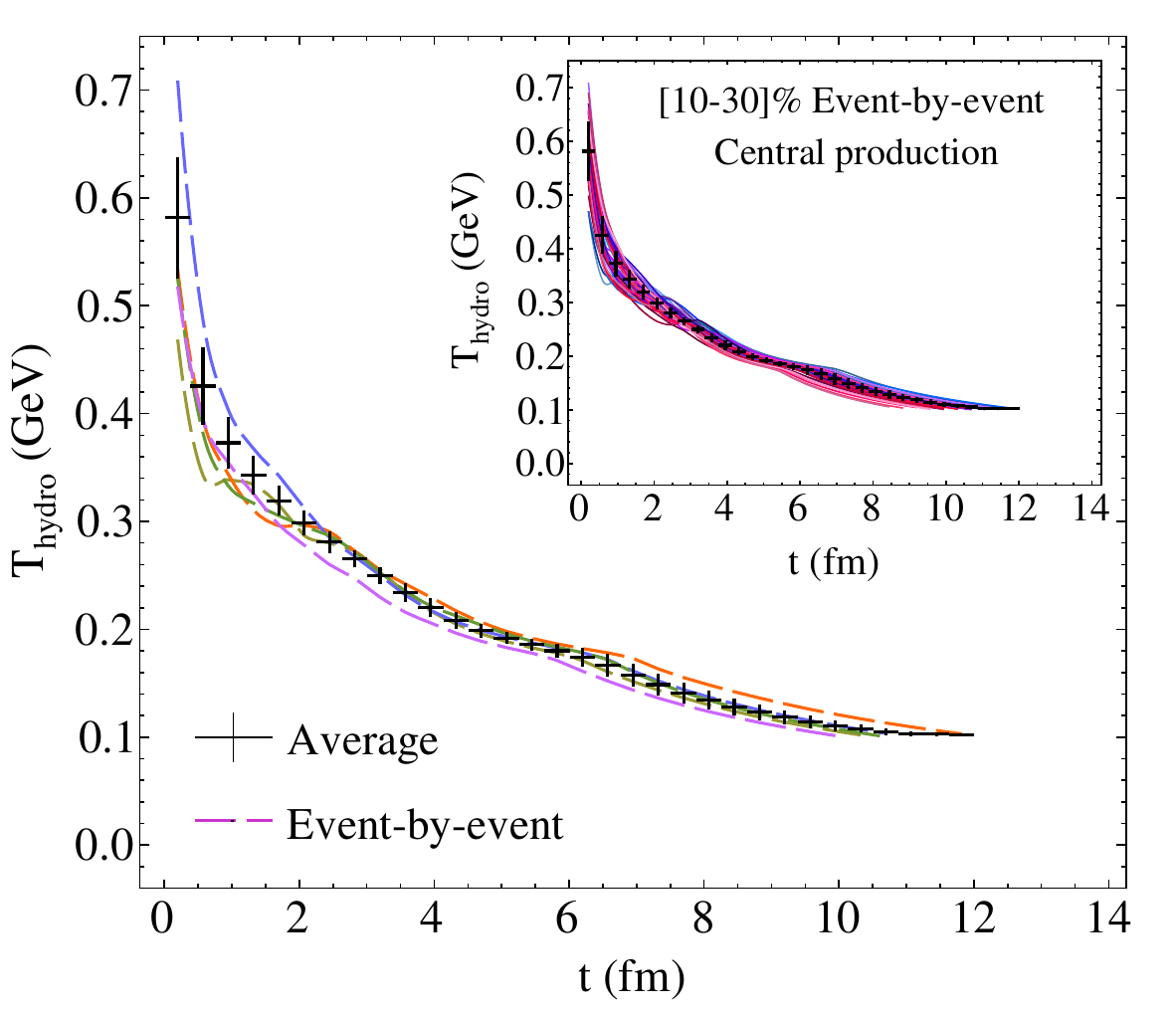}
\caption{Left panel: average temperature profile and corresponding standard deviation (black points) over the 0-10$\%$ most central events in $\sqrt{s_{\rm NN}} =2.76$ TeV Pb-Pb collisions for a parton produced away from the midpoint between the centers of the nuclei and propagating along the the in-plane direction. The dashed colored curves correspond to a selection of representative trajectories employed in the computation of this average. The inset panel shows all trajectories used to obtain the average. Right panel: average temperature profile and corresponding standard deviation (black points) over 10-30$\%$ central events in $\sqrt{s_{\rm NN}} =2.76$ TeV Pb-Pb collisions) for a parton produced at midpoint between the centers of the nuclei and propagating with a $210^\circ$ angle with respect to the in-plane direction. The dashed colored curves correspond to a selection of representative trajectories employed in the computation of this average. The inset panel shows all trajectories used to obtain the average}
\label{fig:ebye_average}
\end{figure}

To examine the impact of fluctuations along an in-medium path, we focus on two centrality classes (0-10$\%$ and 10-30$\%$) selected from a total of 400 minimum-bias $\sqrt{s_{\rm NN}} = 2.76$ TeV Pb-Pb events. We then select straight-line trajectories with the same production point and direction over all the events in a given centrality. This setup allow us to clearly evaluate the fluctuations induced from an event-by-event analysis on our proposed power-law scaling. For the 0-10$\%$ centrality class, we select the trajectory of a hard parton produced at $\tau_{\rm hydro}$ 4 fm away both in the $x-$ and $y$-directions from the midpoint between the centers of the two nuclei (off-central production), and propagating along the in-plane direction. For the 10-30$\%$ centrality class, the hard parton is produced at the midpoint between the centers of the two lead nuclei (central production), and  propagates with a $210^\circ$ with respect to the in-plane direction. The full list of obtained trajectories is shown as coloured lines in the inset panels of figure~\ref{fig:ebye_average}. The resulting average temperature profile obtained from all the sampled trajectories within the chosen centrality class is shown in figure~\ref{fig:ebye_average} as black points, together with its respective standard deviation illustrated by the vertical lines. To provide a clearer visualization, we also show a selection of temperature profiles (colored dashed lines) sampled from the total list of selected trajectories shown in the inset panels of figure~\ref{fig:ebye_average}. 

As it can be seen on the left panel of figure~\ref{fig:ebye_average}, some of the resulting off-central temperature profiles do not monotonically decrease with time. Furthermore, pronounced fluctuations with respect to the average are observed. Conversely, trajectories sampled from the 10-30$\%$ centrality  display a more monotonic behavior, albeit with a steeper temperature evolution, as it can be seen in the right panel of the same figure. In both of these scenarios, our scaling law may deviate from its highest accuracy. By properly evaluating these contrasting scenarios, we can establish an upper limit on the potential error incurred when using our proposed power-law scaling instead of employing the full numerical solution along  each trajectory.

\begin{figure}
\includegraphics[width=\textwidth]{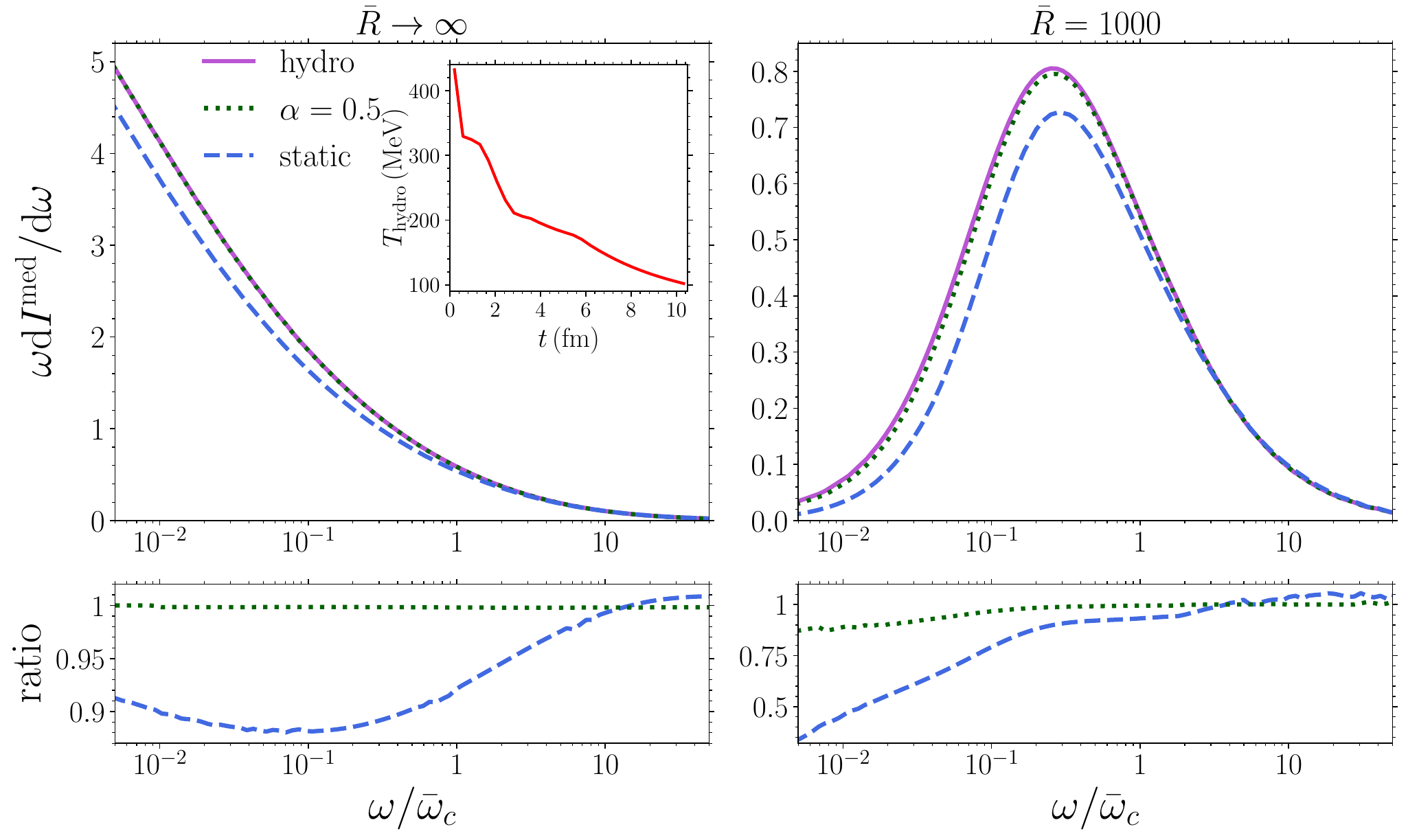}
\caption{Top: fully resummed energy distribution for the Yukawa-type interaction with $\chi = 5$, and $ \bar R \rightarrow \infty$ (left panel) or $ \bar R  = 1000$ (right panel) as a function of $\omega/\bar \omega_c$. The purple curves correspond to the spectra along the temperature profile shown in the inset figure sampled from an event corresponding to the 10-30$\%$ centrality class in $\sqrt{s_{\rm NN}} =2.76$ TeV Pb-Pb collisions at the LHC. For detailed information about the employed path, see main text. The dashed blue and dotted green curves correspond, respectively, to the results for the static ((see eqs.~\eqref{eq:1moment_static}-\eqref{eq:2moment_static}) ) and power-law scenarios (see eqs.~\eqref{eq:0moment_powerlaw}-\eqref{eq:2moment_powerlaw}). Bottom: ratio of the power-law (static) spectrum w.r.t. the spectrum along the path in dotted green (dashed blue).}
\label{fig:LHC_ebye}
\end{figure}

In the following, we select different events out of the previous samples. 
In figure~\ref{fig:LHC_ebye}, we present
the energy spectrum computed along the path shown in its inset panel, which was selected from the paths shown on the right panel of figure~\ref{fig:ebye_average} (10-30$\%$ centrality). This temperature profile displays significant fluctuations with respect to the average shown in the right panel of figure~\ref{fig:ebye_average}. The spectrum along this path is further compared to the static scenario obtained through the scaling laws \eqref{eq:1moment_static}-\eqref{eq:2moment_static} (dashed blue curve) and the power-law equivalent medium given by \eqref{eq:0moment_powerlaw}-\eqref{eq:2moment_powerlaw} (dotted green curve). In both panels, we set $k_1=0.5$  (or equivalently $\chi=5$), with the left panel corresponding to the $\bar R \rightarrow \infty$ limit and the right panel to $\bar R=1000$ (or, equivalently, $k_2=50$).  Despite the fluctuations being clearly visible along the considered path, we observe that the power-law scaling yields an accurate description of the spectrum for all gluon energies regardless of whether the kinematic constraint is removed (left panel) or imposed (right panel).

\begin{figure}
\includegraphics[width=\textwidth]{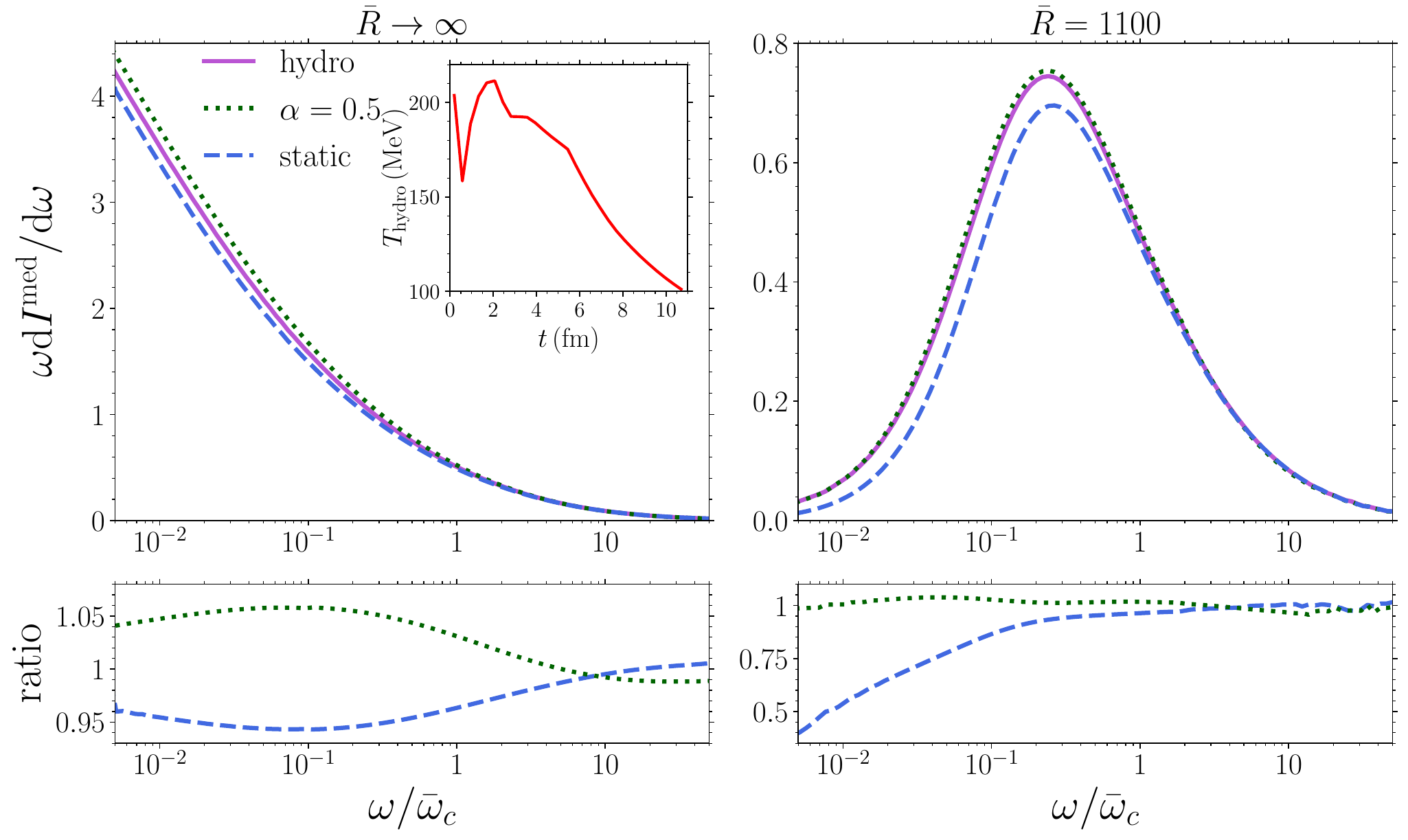}
\caption{Top: fully resummed medium-induced energy distribution for the Yukawa interaction model with $\chi = 4.4$, and $ \bar R \rightarrow \infty$ (left panel) or $ \bar R  = 1100$ (right panel) as a function of $\omega/\bar \omega_c$. The purple curves correspond to the spectra along the temperature profile shown in the inset figure sampled over a central event in $\sqrt{s_{\rm NN}} =2.76$ TeV Pb-Pb collisions at the LHC. For detailed information about the employed path, see main text. The dashed blue and dotted green curves correspond, respectively, to the results for the static (see eqs.~\eqref{eq:1moment_static}-\eqref{eq:2moment_static})  and power-law scenarios (see eqs.~\eqref{eq:0moment_powerlaw}-\eqref{eq:2moment_powerlaw}). Bottom: ratio of the power-law (static) spectrum w.r.t. the spectrum along the path in dotted green (dashed blue).}
\label{fig:LHC_ebye_extreme}
\end{figure}

An example of the energy spectrum computed along a fluctuating and non-monotonic in-medium path is shown in figure~\ref{fig:LHC_ebye_extreme} (solid purple curve). The inset panel illustrates the corresponding temperature profile selected from the paths shown in the left panel of figure~\ref{fig:ebye_average} (0-10$\%$ centrality). We employ in this figure the same values of $k_1=0.5$ and $k_2=50$ as in the previous one, resulting for this path in $\chi=4.4$ and $\bar R\rightarrow \infty$ (left panel) and $\chi=4.4$ and $\bar R=1100$ (right panel). We observe that even in such an extreme scenario, the power-law scaling (dotted green line) yields an excellent description of the spectrum computed along the actual path across its full kinematic range, both when the kinematic constraint is imposed or removed, while the static matching (dashed blue line) only provides a suitable description in the high-energy tail of the spectrum.

Overall, these two examples illustrate that even when event-by-event fluctuations are taken into account, the in-medium gluon energy spectrum can be well approximated with the simple power-law matching proposed in section~\ref{subsec:power}. Finally, we refer the reader to appendix~\ref{sec:appHTL}, where we show that the power-law scaling works remarkably well when considering other interaction models, such as the HTL collision rate. All together, these results highlight the effectiveness of the power-law scaling approach in approximating the spectrum along realistic hydrodynamic paths for a wide range of trajectories and centrality classes in Pb-Pb collisions at the LHC.

\section{Conclusions and outlook}
\label{sec:conclusions}

In this manuscript, we  advance in the study of medium-induced gluon radiation by including the effects of the longitudinal expansion of the medium. By employing a framework that incorporates full resummation of multiple scatterings  \cite{Andres:2020vxs,Andres:2020kfg}, we are able to compute the in-medium emission spectrum for any time variation of its parameters. This represents a substantial improvement over previous approaches, such as the Harmonic Oscillator approximation, which limited the evaluation of the spectrum to specific functional forms of the time dependence \cite{Arnold:2008iy}. 

Our method not only allows us to compute the spectrum for realistic conditions with trajectories extracted from hydrodynamical simulations, but it also provides a quantitative assessment of the accuracy of scaling laws used to assign an ``equivalent static scenario'' to arbitrary trajectories. \cite{Salgado:2002cd,Salgado:2003gb}. We have found that, as expected, the equivalent static result closely resembles the spectrum obtained from the exact evaluation of the hydrodynamically evolving profile for large values of the gluon energy $\omega$, as demonstrated in section~\ref{subsec:high-energy-tail}. However, significant differences of up to $50\%$ arise in the low-energy region of the spectrum. These disparities can be attributed to the fact that softer gluons have considerably shorter formation times, rendering them more sensitive to the early dynamics when the evolving medium is hotter, a characteristic absent in the static case.

Given that our current method allows the evaluation of the in-medium emission spectrum for any input evolution of the parameters, the reliance on scaling laws may appear redundant, especially considering the significant errors they introduce in certain regions of phase space. Nevertheless, the ability to approximate the in-medium spectrum for an arbitrary trajectory with an equivalent pre-evaluated and tabulated case holds immense applicability in phenomenological studies.
In such analyses,  it is often necessary to compute the emission spectrum for a large number of trajectories extracted from numerous events. As a result, the computational cost of evaluating the spectrum for each individual trajectory becomes a significant concern. In order to address this issue, having a pre-evaluated set of spectra along with their corresponding scaling laws is still highly desirable.  With this motivation, we have proposed the use of \emph{power-law scaling laws} to relate any trajectory to a power-law decrease in the medium parameters, rather than relying on the static case.

Our results demonstrate the remarkable accuracy of the newly proposed scaling law in describing the energy spectrum across a wide range of realistic in-medium straight-line paths. These trajectories were sampled from 2+1 viscous hydrodynamic simulations of the QGP generated in various centrality classes in $\sqrt{s_{\rm NN}} = 5.02~\rm{TeV}$ Pb-Pb collisions at the LHC. Even when considering event-by-event fluctuations (see section~\ref{sec:ebye}) and off-central production points (see figure~\ref{fig:LHC2_b3.2_x0y80theta255}), the deviations between the spectrum along the path and its power-law equivalent remained consistently below 15$\%$ across all gluon energies. Furthermore, we have tested the accuracy of the power-law scaling for the collision rate derived from HTL calculations, and found a substantial improvement compared to previous static-equivalent matching scenarios (see appendix~\ref{sec:appHTL}). In conclusion, our analysis shows that a medium characterized by a power-law decay with a power of $0.5$ and an initial offset relative to the medium length of $0.1$ provides the best approximation to the energy spectrum computed along realistic temperature profiles. 

Having upgraded the formalism of \cite{Andres:2020vxs,Andres:2020kfg} to account for longitudinal expansion, it becomes feasible to go one step further and also upgrade the phenomenological studies performed in \cite{Andres:2016iys,Andres:2019eus} and evaluate the effect of using a more realistic parton-medium interaction in the energy loss calculations. Several intermediate steps would be needed to achieve this objective, most notably the computation of the quenching weights \cite{Salgado:2003gb} which account for the energy loss due to multiple independent emissions. These studies will be left for future publications, where the effects of the initial stages after the collision should also be considered, as done in \cite{Andres:2022bql}.

It is also important to acknowledge the existence of other sources of error which might impact phenomenological studies using this formalism. In particular, two approximations taken in this approach  may need to be revisited in the near future: the high-energy approximation which neglects contributions suppressed by powers of the initial parton energy, and the fully coherent approximation, which assumes that the jet loses energy as a single source without accounting for differences in the total amount of radiation emitted by a jet due to fluctuations in its branching. Contributions suppressed by powers of the initial energy include the effects of considering flow or density gradients as shown in \cite{Sadofyev:2021ohn,Barata:2022krd,Andres:2022ndd,Barata:2023qds,Fu:2022idl}. These are expected to be small for most observables, but quantitative estimates of the size of the error incurred by neglecting them will only be available once the theoretical tools currently being developed reach the level of accuracy of the techniques used in this paper. Regarding the fully coherent approximation, some theoretical developments attempting to account for differences in the energy lost by a jet due to its substructure are available for the static brick case \cite{Mehtar-Tani:2017ypq,Mehtar-Tani:2017web}, with those results being used in recent phenomenological studies \cite{Mehtar-Tani:2021fud,Mehtar-Tani:2024jtd}. Nevertheless, a detailed discussion of how these advancements must be accurately and effectively implemented in expanding media is still lacking. In a more general context, it is important to highlight that jet-by-jet fluctuations have been studied in numerous works, where they have been shown to play an important role in the description of many jet observables, as noticed first for the dijet asymmetry \cite{Milhano:2015mng,Escobedo:2016vba,Chang:2016gjp}, and subsequently extended to many other cases \cite{Escobedo:2016jbm,Rajagopal:2016uip,Tachibana:2017syd,Chang:2017gkt,Casalderrey-Solana:2018wrw,Brewer:2018mpk,He:2018xjv,Casalderrey-Solana:2019ubu,Caucal:2019uvr,Brewer:2021hmh}.


\acknowledgments

We thank Carlos A. Salgado for insightful discussions about this work and for carefully reading this manuscript. This work is supported by European Research Council project ERC-2018-ADG-835105 YoctoLHC; by Maria de Maetzu excellence program under project CEX2020-001035-M; by Spanish Research State Agency under project PID2020-119632GB-I00; by OE Portugal, Funda\c{c}\~{a}o para a Ci\^{e}ncia e a Tecnologia (FCT), I.P., projects EXPL/FIS-PAR/0905/2021 and CERN/FIS-PAR/0032/2021; by European Union ERDF. This work has received financial support from Xunta de Galicia (CIGUS Network of Research Centers). C.A. has received funding from the European Union’s Horizon 2020 research and innovation program under the Marie Sklodowska-Curie grant agreement No 893021 (JQ4LHC). L.A. was supported by FCT under contract 2021.03209.CEECIND. M.G.M. was supported by Ministerio de Universidades of Spain through the National Program FPU (grant number FPU18/01966).

\appendix
\section{Dependence of the power-law scaling on $\alpha$ and $t_0$}
\label{sec:appA}

In this appendix we present additional figures showing the dependence of the power-law scaling introduced in section~\ref{subsec:power} on the value of the power-law parameters $\alpha$ and $t_0=\bar t_0/\bar L$ (see eq.~\eqref{eq:powerlaw}).

\begin{figure}
\includegraphics[width=\textwidth]{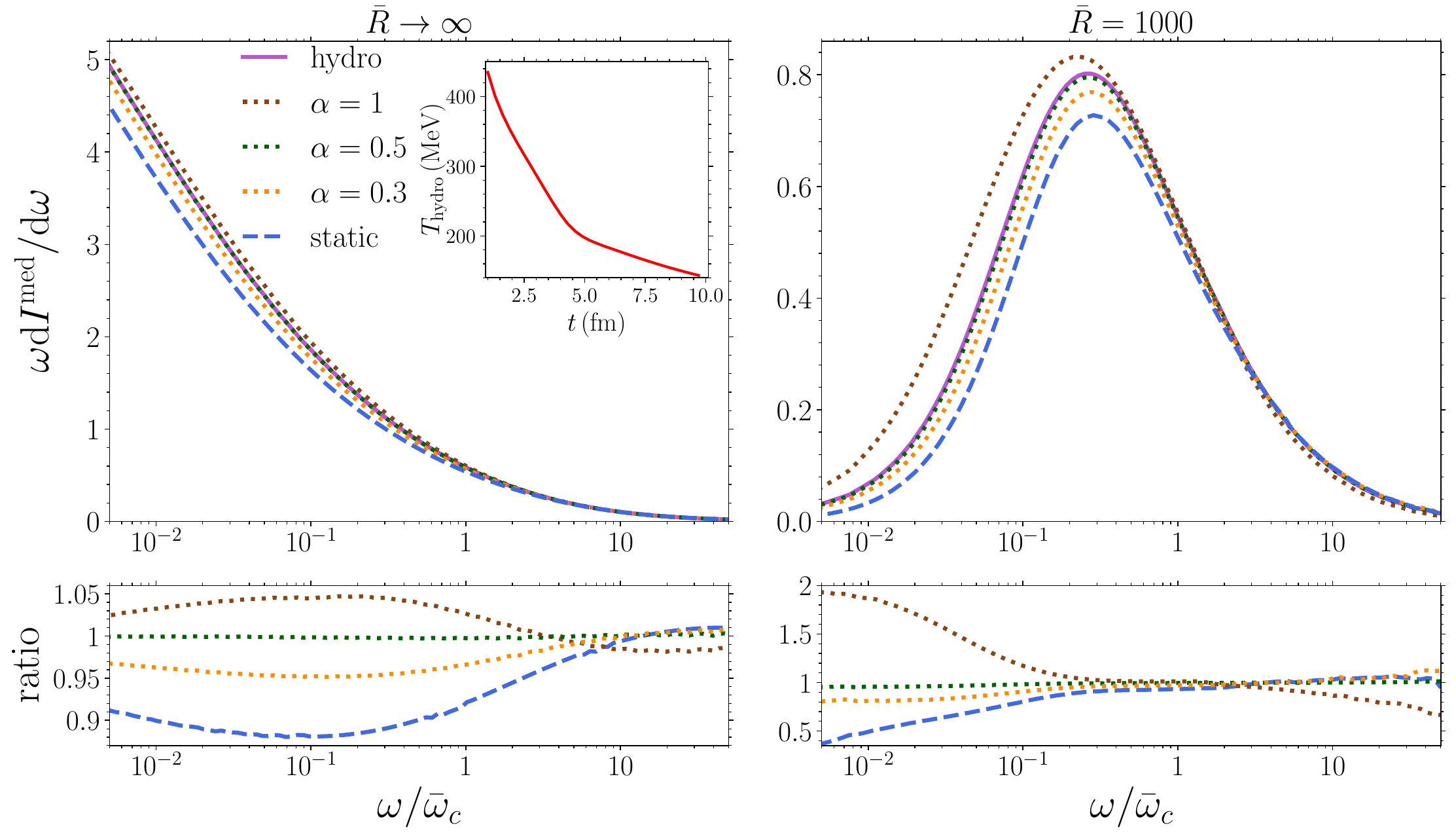}
\caption{Top: fully resummed medium-induced energy distribution for the Yukawa-type interaction with $\chi = 5$, and $ \bar R \rightarrow \infty$ (left panel) or $ \bar R=1000$ (right panel) as a function of $\omega/\bar \omega_c$. The purple curves correspond to the spectra along the temperature profile shown in the inset panel sampled over the 0-10$\%$ centrality class in $\sqrt{s_{\rm NN}} =5.02$ TeV Pb-Pb collisions at the LHC. For detailed information about the employed path, see main text. The dashed blue curve refer to the results for the static (see eqs.~\eqref{eq:1moment_static}-\eqref{eq:2moment_static}) scenario. The dotted curves correspond to the power-law matching (see eqs.~\eqref{eq:0moment_powerlaw}-\eqref{eq:2moment_powerlaw}) with $t_0=0.1$ and $\alpha=0.3$ (orange), $\alpha=0.5$ (green), and $\alpha=1.0$ (brown). Bottom: ratio of the power-law and static spectra w.r.t. the spectrum along the temperature profile.}
\label{fig:LHC2_b3.2_x0y0theta0_alphas}
\end{figure}

\begin{figure}
\includegraphics[width=\textwidth]{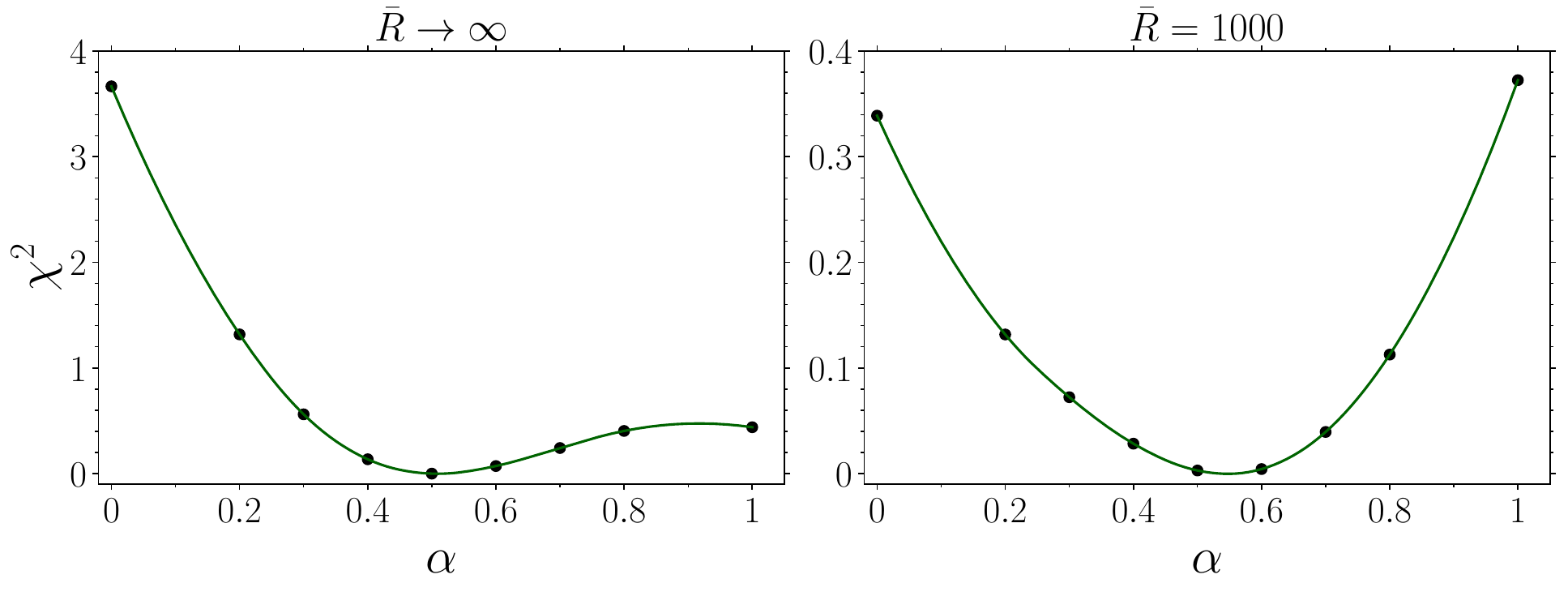}
\caption{$\chi^2$-values for different values of $\alpha$ for the spectrum along the hydrodynamic path and medium parameters in the previous figure. The left panel corresponds to  $\bar R \rightarrow \infty$ and the right one to $ \bar R=1000$. }
\label{fig:chi2}
\end{figure}

Firstly, we show in figure~\ref{fig:LHC2_b3.2_x0y0theta0_alphas} a comparison between spectrum computed along a typical path sampled over the 0-10$\%$ centrality class in $\sqrt{s_{\rm NN}} =5.02$ TeV Pb-Pb collisions at the LHC, and the power-law matched result described in section~\ref{subsec:power} for several values of $\alpha$ (and $t_0$ fixed to $t_0=0.1$). We use in this figure the same parameters values $k_1=0.5$ and $k_2=45$ and the same trajectory, shown in the inset panel, as those employed in figure~\ref{fig:LHC2_b3.2_x0y0theta0_a05}, corresponding to a hard parton produced at the midpoint between the centers of the two lead nuclei and moving along the in-plane direction. It is evident that the power-law with $\alpha=0.5$ (dotted green curve) provides the best description of the spectrum along this path (solid purple curve), regardless of whether the kinematic constraint is lifted (left panel) or imposed (right panel). Additionally, we have performed a $\chi^2$ fit to determine the value of $\alpha$ that best describes the spectrum along the path with the kinematic constraint imposed (lifted) in the right (left) panel of figure~\ref{fig:LHC2_b3.2_x0y0theta0_a05}. The values of the $\chi^2$ for different values of $\alpha$ are presented in figure~\ref{fig:chi2}. We note that additional values of $\alpha$ have been incorporated in this figure, which were not shown in the previous one to not difficult its visibility. The minimum is located at $\alpha=0.51$ when the kinematic cutoff is removed (left panel) and at $\alpha=0.54$ both when the kinematic cutoff is included (right panel).

Upon further examination of additional trajectories, over several centrality classes, and with varying values for the $k_1$ and $k_2$ parameters, we consistently found the minima of the $\chi^2$-fit to occur at $\alpha \simeq 0.5$. Given that the power-law scaling consistently and accurately characterizes the spectrum along the actual hydrodynamics paths at this specific $\alpha$ value, we have opted to fix $\alpha= 0.5$ instead of treating it as a free parameter. For completeness, we provide in tables~\ref{tab:chi2_rinf}~and~\ref{tab:chi2_rfin} the values of the $\chi^2$ as function of $\alpha$ corresponding to the spectra presented in all figures in section~\ref{sec:matching}. For the spectra shown in the left panel (without kinematical constraint) figure~\ref{fig:LHC2_b3.2_x0y80theta255}, we find that the minimum of the $\chi^2$ fit is located at a slightly smaller value of alpha, although that is not the case for the spectra in the right panel (with kinematical constraint). This discrepancy is not an issue, as trajectories with a non-monotonically decreasing temperature profile are uncommon and carry little weight in the computation of any observable. Moreover, since it only arises in the case with no kinematical constraint, it will have no impact on phenomenological applications in which the spectrum with kinematic constraint is employed.

\begin{table}
    \centering
    \begin{tabular}{c|c|c|c|c|c}
     & $\alpha=0$ & $\alpha=0.3$ & $\alpha=0.5$ & $\alpha=0.7$& $\alpha=1$ \\
     \hline
    Figures~\ref{fig:LHC2_b3.2_x0y0theta0_a05}~and~\ref{fig:LHC2_b3.2_x0y0theta0_alphas}  &3.67  & 0.56 & 0.0006 & 0.24 &0.44\\
    Figure~\ref{fig:LHC2_b3.2_x10y10theta90_a05}    & 1.49 & 0.18  &  0.001& 0.12 & 0.16\\
    Figure~\ref{fig:LHC2_b7.2_x0y30theta270}     & 7.51  & 0.81 & 0.067  &1.14 &2.06 \\
    Figure~\ref{fig:LHC2_b3.2_x0y80theta255}    & 4.71 &  0.06 & 1.00 &3.49 & 5.32 \\      
    \end{tabular}
    \caption{$\chi^2$-values for different values of $\alpha$ for the spectra without kinematic constraint shown in left panels of the figures in section~\ref{sec:matching}.}
    \label{tab:chi2_rinf}
\end{table}

\begin{table}
    \centering
    \begin{tabular}{c|c|c|c|c|c}
     & $\alpha=0$ & $\alpha=0.3$ & $\alpha=0.5$ & $\alpha=0.7$& $\alpha=1$ \\
     \hline
Figures~\ref{fig:LHC2_b3.2_x0y0theta0_a05}~and~\ref{fig:LHC2_b3.2_x0y0theta0_alphas}  & 0.34 & 0.07  & 0.003 & 0.04 & 0.37\\
    Figure~\ref{fig:LHC2_b3.2_x10y10theta90_a05} & 0.071  &  0.013&  0.0001&  0.014&   0.11\\
    Figure~\ref{fig:LHC2_b7.2_x0y30theta270}     &0.83  & 0.16  &  0.002 & 0.14 & 1.08\\
    Figure~\ref{fig:LHC2_b3.2_x0y80theta255}    & 0.50 & 0.06 & 0.009  & 0.19 &1.01\\      
    \end{tabular}
    \caption{$\chi^2$-values for different values of $\alpha$ for the spectra with kinematic constraint shown in the right panels of the figures in section~\ref{sec:matching}.}
    \label{tab:chi2_rfin}
\end{table}

\begin{figure}
\includegraphics[width=\textwidth]{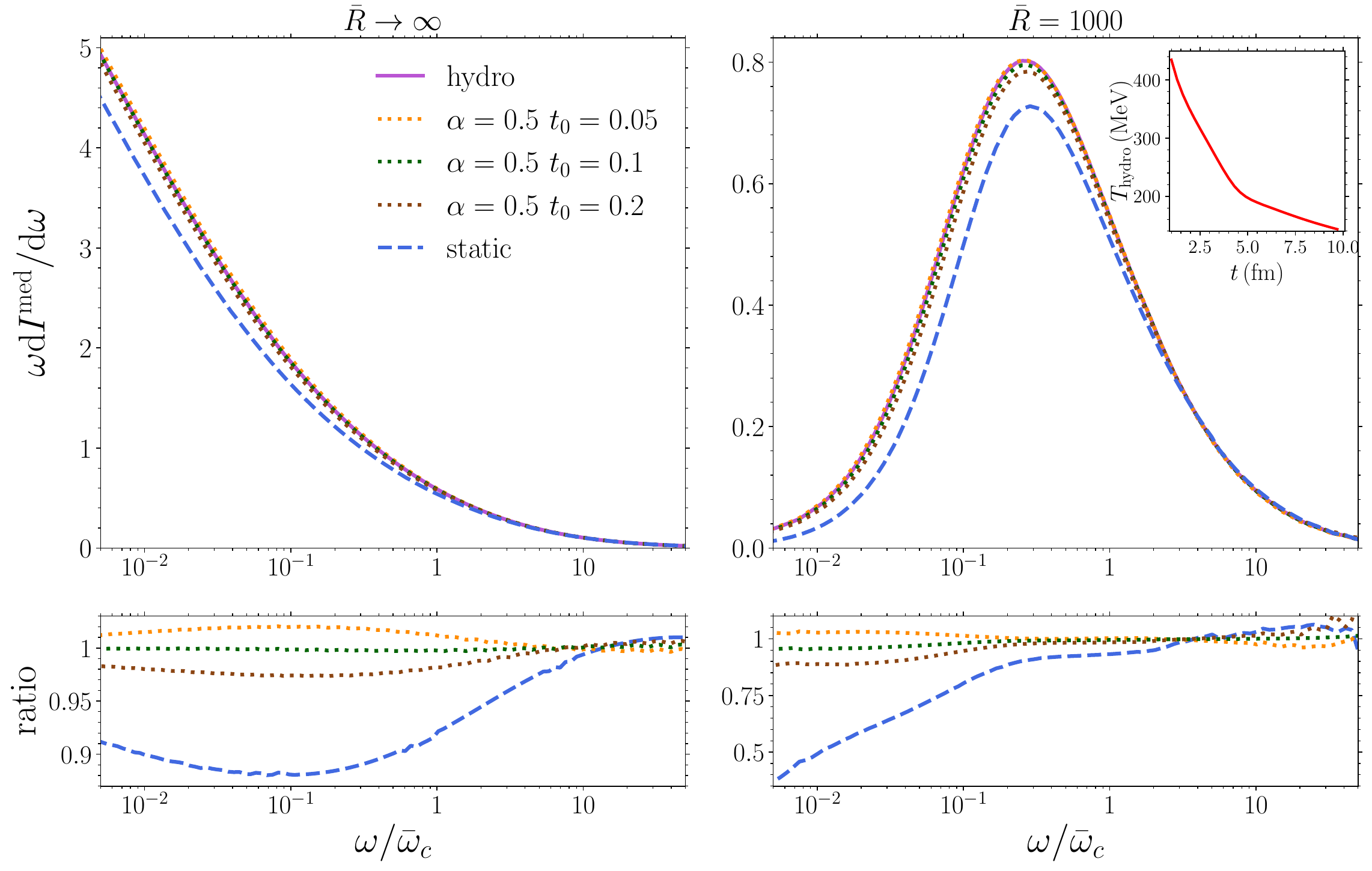}
\caption{Top: fully resummed medium-induced energy distribution for the Yukawa interaction model with $\chi = 5$, and $ \bar R \rightarrow \infty$ (left panel) or $ \bar R  = 1000$ (right panel) as a function of $\omega/\bar \omega_c$. The purple curves correspond to the spectra along the temperature profile shown in the inset figure sampled over the 0-10$\%$ centrality class in $\sqrt{s_{\rm NN}} =5.02$ TeV Pb-Pb collisions at the LHC. For detailed information about the employed path, see main text. The dashed blue refers to the results for the static (see eqs.~\eqref{eq:1moment_static}-\eqref{eq:2moment_static}) scenario. The dotted curves correspond to the power-law scenario  (see eqs.~\eqref{eq:0moment_powerlaw}-\eqref{eq:2moment_powerlaw}) with $\alpha=0.5$ and $t_0=0.05$ (orange), $t_0=0.1$ (green), and $t_0=0.2$ (brown). Bottom: ratio of the power-law and static spectra w.r.t. the spectrum along the temperature profile.}
\label{fig:LHC2_b3.2_x0y0theta0_t0s}
\end{figure}

We further illustrate the dependence of the power-law matching on $t_0$ in figure~\ref{fig:LHC2_b3.2_x0y0theta0_t0s}. We make use of the same trajectory and parameter values $k_1=0.5$ and $k_2=45$  as in the previous figure. Our findings indicate that for $\alpha=0.5$, changing $t_0$ has minimal impact on the resulting power-law spectrum. As a result, we have decided to set $t_0=0.1$ as a fixed value, thus reducing the number of parameters required to employ the power-law scaling approach. Importantly, this selection does not significantly impact the accuracy of the power-law matching, ensuring its reliability in describing the spectrum along realistic hydrodynamic paths at LHC energies.

By fixing $\alpha=0.5$ and $t_0=0.1$, the number of parameters required to determine the power-law equivalent medium that correctly describes the spectrum computed along the hydrodynamic path is reduced to three. These three parameters can be determined through the scaling laws given by eqs.~\eqref{eq:0moment_powerlaw}-\eqref{eq:2moment_powerlaw}. This reduction in parameters will greatly simplify future phenomenological studies while still maintaining the accuracy of the power-law matching approach.

\section{KLN initial conditions}
\label{sec:appKLN}

In this appendix, we analyze the performance of the power-law matching relations described in section~\ref{subsec:power} within a smooth averaged hydrodynamic simulation whose initial condition, unlike the Glauber one used in section~\ref{sec:matching}, depends also on the number of participants.\footnote{As done in section~\ref{sec:matching}, the dimensionless parameters entering the power law \eqref{eq:powerlaw} are fixed to $\alpha=0.5$ and $t_0=0.1$, since they provide the best approximation to the spectrum along the actual hydrodynamic path.}  Specifically, we make use of the 2+1 viscous hydrodynamic model in refs.~\cite{Luzum:2008cw,Luzum:2009sb} taking its initial condition from the factorized Kharzeev-Levin-Nardi (KLN) model \cite{Drescher:2006pi} and with shear viscosity to entropy density ratio fixed to $\eta/s = 0.16$. The simulation, as the Glauber one described in section~\ref{sec:matching}, starts at an initial time $\tau_{\rm hydro}=1$ fm,  uses an equation of state
inspired by Lattice QCD, and assumes the system to be in chemical equilibrium until it reaches a freeze-out temperature $T_{\rm f} = 140$ MeV. Since the primary distinction between this hydrodynamic simulation and the one discussed in section~\ref{sec:matching} lies in their respective initial conditions,  we will denote the hydrodynamics employed in this appendix as KLN.

\begin{figure}
\includegraphics[width=\textwidth]{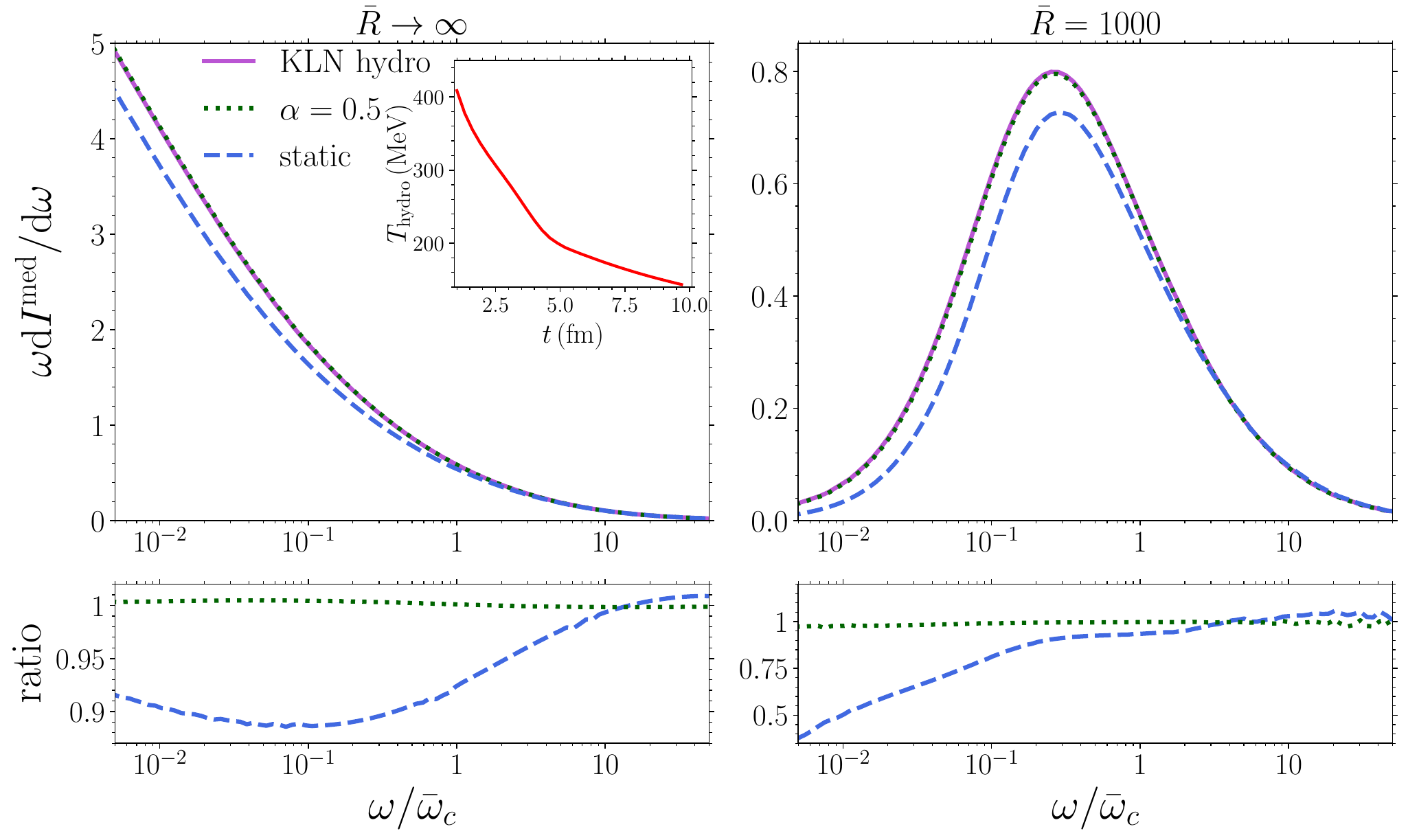}
\caption{Top: fully resummed medium-induced gluon energy distribution for a Yukawa interaction model with $\chi = 5$, and $ \bar R \rightarrow \infty$ (left panel) or $ \bar R  = 1000$ (right panel) as a function of $\omega/\bar \omega_c$. The purple curves correspond to the spectra along the temperature profile shown in the inset figure sampled with a central production point over the 0-10$\%$ centrality class in $\sqrt{s_{\rm NN}} =5.02$ TeV Pb-Pb collisions simulated with the KLN hydrodynamics. For detailed information about the employed path, see main text. The dashed blue and dotted green curves correspond, respectively, to the results for the static (see eqs.~\eqref{eq:1moment_static}-\eqref{eq:2moment_static}) and power-law scenarios (see eqs.~\eqref{eq:0moment_powerlaw}-\eqref{eq:2moment_powerlaw}). Bottom: Ratio of the power-law (static) spectrum w.r.t. the spectrum along the temperature profile in dotted green (dashed blue).}
  \label{fig:LHC2_fKLNb3.2_x0y0theta0_a05}
\end{figure}

We present in figure~\ref{fig:LHC2_fKLNb3.2_x0y0theta0_a05} the medium-induced spectrum for a trajectory that follows the same path used in figures~\ref{fig:LHC2_b3.2_x0y0theta0_static}~and~\ref{fig:LHC2_b3.2_x0y0theta0_a05}, i.e. a straight-line starting at the mid-point between the centers of the two lead nuclei and going in the in-plane direction. In this figure, this path is sampled from the KLN hydrodynamic simulation of 0-10$\%$ Pb-Pb  collisions at $\sqrt{s_{\rm NN}} =5.02$ TeV, instead of the Glauber one used in figures~\ref{fig:LHC2_b3.2_x0y0theta0_static}~and~\ref{fig:LHC2_b3.2_x0y0theta0_a05}. As it can be seen in the inset panel of figure~\ref{fig:LHC2_fKLNb3.2_x0y0theta0_a05}, the resulting temperature profile along the KLN hydrodynamics is, as expected, different to that of the Glauber simulation shown in the inset panels of figures~\ref{fig:LHC2_b3.2_x0y0theta0_static}~and~\ref{fig:LHC2_b3.2_x0y0theta0_a05}. Regarding the spectrum, we use the same parameters as figures~\ref{fig:LHC2_b3.2_x0y0theta0_static}~and~\ref{fig:LHC2_b3.2_x0y0theta0_a05}, i.e, $\chi = 5$, with the left panel corresponding to  $\bar R \rightarrow \infty $  and the right panel to  $\bar R = 1000$. We compare the spectrum along the KLN path (solid purple curve) to the static scenario though the matching relations \eqref{eq:1moment_static}-\eqref{eq:2moment_static} (dashed blue curve) and the proposed power-law scaling given by eqs.~\eqref{eq:0moment_powerlaw}-\eqref{eq:2moment_powerlaw}  (dotted green curve). As for the Glauber initial conditions, the power-law scaling yields an accurate description of the spectrum along the KLN hydrodynamic path for all gluon energies, regardless of whether the kinematic constraint is removed (left panel) or imposed (right panel). It is worth noting that, akin to the Glauber case in figure~\ref{fig:LHC2_b3.2_x0y0theta0_a05}, the power-law scaling outperforms the static approach along the full kinematic range.

We have further conducted an extensive analysis of the emission spectra along straight-line trajectories sampled from the KLN hydrodynamic simulation of $\sqrt{s_{\rm NN}} = 5.02$ TeV Pb-Pb collisions across various centrality classes and a wide range of medium parameters. Our findings reveal that, for monotonically decreasing temperature profiles, such as the one depicted in figure~\ref{fig:LHC2_fKLNb3.2_x0y0theta0_a05}, the power-law matching approach offers a highly accurate description of the spectrum along the entire gluon kinematics. Specifically, the ratio between the power-law spectrum and the actual spectrum along the path consistently remains below 1\% for all monotonically decreasing temperature profiles sampled over central to semi-peripheral LHC Pb-Pb collisions within the KLN hydrodynamics. To avoid unnecessary lengthening of the manuscript, we refrain from including additional figures illustrating these results here, as they closely resemble the Glauber outcomes presented in section~\ref{sec:matching}. 

\begin{figure}
\includegraphics[width=\textwidth]{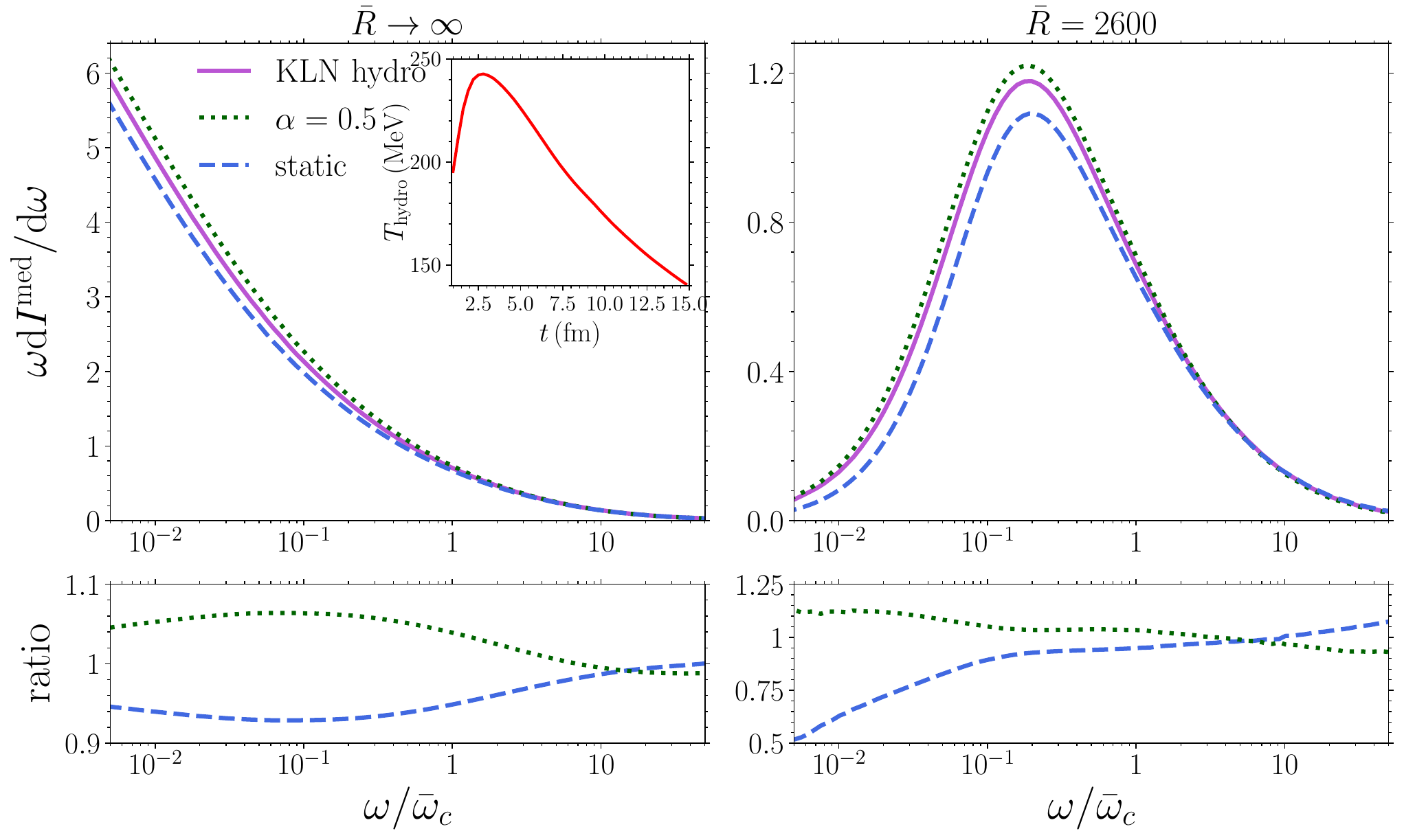}
\caption{ Top: fully resummed medium-induced energy distribution for the Yukawa interaction model with  $\chi=6.6$, and $ \bar R  \rightarrow \infty$ (left panel) or $ \bar R  = 2600$ (right panel) as a function of $\omega/\bar{\omega}_c$. The purple curves correspond to the spectra along the rare temperature profile shown in the inset figure sampled with a non-central production point over the 0-10$\%$ centrality class in $\sqrt{s_{\rm NN}} =5.02$ TeV Pb-Pb collisions simulated with the KLN hydrodynamics. For detailed information about the employed path, see main text. The dashed blue and dotted green curves correspond, respectively, to the spectra for the static (see eqs.~\eqref{eq:1moment_static}-\eqref{eq:2moment_static}) and power-law matchings (see eqs.~\eqref{eq:0moment_powerlaw}-\eqref{eq:2moment_powerlaw}). Bottom: ratio of the power-law (static) spectrum w.r.t. the spectrum along the temperature profile in dotted green (dashed blue).}
\label{fig:LHC2_KLNb3.2_x0y80theta255}
\end{figure}

For (rare) temperature profiles that do not monotonically decrease with time, sampled over the KLN hydrodynamics, the power-law scaling consistently proves to be a superior approximation of the spectrum along the actual path, with discrepancies always below $15\%$, compared to the static scaling.  This is illustrated in figure \ref{fig:LHC2_KLNb3.2_x0y80theta255}
for a trajectory sampled over the 0-10$\%$ centrality class in $\sqrt{s_{\rm NN}} =5.02$ TeV Pb-Pb collisions simulated with the KLN hydrodynamics,
where the temperature profile exhibits the non-monotonically decreasing behavior shown in the inset panel. This trajectory as the one in figure~\ref{fig:LHC2_b3.2_x0y80theta255} corresponds to a hard parton produced in the edge of overlapping area between the nuclei. Specifically, the production point is located in the out-of-plane direction at a distance of 8\,fm form the mid-point between the centers of the two lead nuclei and the parton propagates inwards with an angle of 255$^{\circ}$ with respect to the in-plane direction. Comparing the inset panels of figures~\ref{fig:LHC2_b3.2_x0y80theta255} and~\ref{fig:LHC2_KLNb3.2_x0y80theta255}, we observe differences in the resulting temperature profiles along this path within the two different hydrodynamic implementations. For the medium parameters, we use  $\chi=6.6$, and $\bar R \rightarrow \infty$ (left panel) and $\bar R= 2600$ (right panel). 
Remarkably, despite this highly extreme scenario, the deviations of the power-law spectrum (dotted green) from the actual spectrum along the KLN temperature profile remain below 15\% across the entire range of gluon kinematics. In contrast, the deviations for the static scaling (dashed blue) can reach up to 50\% for low gluon energies when the kinematic constraint is imposed.

In summary, these two examples illustrate that there is no discernible variation in the performance of the power-law matching method proposed in section~\ref{subsec:power} when a different hydrodynamic simulation with a more realistic initial condition, dependent on both the number of binary scatterings and the number of participants, is employed.

\section{Hard thermal loop interaction}
\label{sec:appHTL}

To illustrate the flexibility of our approach, we additionally study the energy spectrum with the collision rate derived from hard thermal loop (HTL) calculations. For this purpose, we employ the leading-order result in thermal field theory for a weakly-coupled medium \cite{Aurenche:2002pd}, which is given by
\beq
\frac{1}{2}n(s)V_{\rm H}(s; \vec q)=\frac{g_s^2 N_c m^2_{\rm D}(s)T(s)}{\vec q^2(\vec q^2 +m^2_{\rm D}(s))}\,,
\label{eq:V_HTL}
\eeq
where $m_{\rm D}(s)$ and $T(s)$ are, respectively, the time-dependent Debye mass and medium temperature. The computation of the fully resummed in-medium radiation spectra in eqs.~\eqref{eq:full_kspec},~{\eqref{eq:fullspec_rfin},~and~\eqref{eq:fullspec_Rinf} using this collision rate was extensively described in \cite{Andres:2020vxs}.

It is worth noting that the collision rate in \eqref{eq:V_HTL} depends only on $m_{\rm D}$ and $T$. Therefore, for a brick of length $L$ (static case), it is evident that the energy distribution \eqref{eq:fullspec_rfin} for this type of parton-medium interaction is function of: $T$, $m_{\rm D}^2$, and $L$. For convenience, we instead adopt the following
\beq
\chi_{\rm H} = TL\,, 
\quad 
\bar \omega_c^{\rm H} =\frac{1}{2}m_{\rm D}^2L\,,
\quad \mathrm{and} \quad
\bar R_{\rm H} = \bar \omega_c^{\rm H} L\,.
\eeq

Moving now to realistic dynamic evolving scenarios, the computation of the all-order spectrum along the emitter's path $\xi(t)$ requires establishing a relation between the Debye mass and the local properties of the medium. At LO in HTL approach, this relation is given by \cite{Aurenche:2002pd}
\beq
m^2_{{\rm D, \,hydro}}(t) = c_2 T_{\rm hydro}^2(\xi(t))\,,
\eeq
where $T_{\rm hydro}$ is the temperature along the trajectory $\xi(t)$ to be extracted from the hydrodynamic simulation in \cite{Luzum:2008cw,Luzum:2009sb}. We note that at LO in the HTL $c_2$ is given by $c_2=g_{\rm s}^2(1+N_f/6)$. For convenience and in analogy to the Yukawa collision rate, we decide to treat here $c_2$ as a free parameter.

\subsection{Scaling laws with respect to static media}

As done for the Yukawa collision rate, we can try to find matching relations for the parameters for the static evaluation that approximate the behavior of the spectrum in an expanding medium. These static scaling laws are given by:
\beq
  \chi_{\rm H} = c_1
 \int_{\tau_{\rm hydro}}^{L + \tau_{\rm hydro}} \mathrm{d}t \,T_{\mathrm{hydro}}(t) \,,
 \label{eq:0moment_static_HTL}
 \eeq
 \beq
\chi_{\rm H} \bar \omega_c^{\rm H}  =
c_1 \int_{\tau_{\rm hydro}}^{L + \tau_{\rm hydro}} {\rm d}t \, t\,T_{\rm hydro}(t) \,m^2_{{\rm D,\,hydro}}(t)\,,
\label{eq:1moment_static_HTL}
 \eeq
 \beq
   \chi_{\rm H} \bar R_{\rm H}
 = \frac{3\,c_1}{2} \int_{\tau_{\rm hydro}}^{L + \tau_{\rm hydro}} \mathrm{d}t \,t^2\,T_{\mathrm{hydro}}(t)\,m^2_{\mathrm{D,\, hydro}}(t) \,,
 \label{eq:2moment_static_HTL}
 \eeq
where we have introduced the constant $c_1$ in analogy to $k_1$ for the Yukawa collision rate, see section~\ref{sec:matching}.

\subsection{Scaling laws with respect to a power-law profile}

As done for the Yukawa parton-medium interaction model in section~\ref{subsec:power}, we now propose to establish an equivalent scenario characterized by a power-law behavior. In the power-law approach, we express the temperature $T$ and the Debye mass $m_{\rm D}$ as follows:
\beq
T(t)=\frac{\bar{T}_0}{(t+\bar{t}_0)^{\alpha}}\,,
\quad 
\mathrm{and}
\quad
m_{\rm D}^2(t)=\frac{\bar{m}_{\rm D \,0}^2}{(t+\bar{t}_0)^{2\alpha}}\,,
\label{eq:powerlawhtl}
\eeq
with $\bar{T}_0$ and  $\bar{m}_{D\,0}$ proportional to the maximum value of the temperature and Debye mass, respectively, at the initial time $\bar{t}_0$. In the following, we will set $\alpha=0.5$ and $t_0=\bar t_0/\bar L=0.1$, as done for the Yukawa collision rate.

The scaling laws for this power-law profile are given by
\beq
 \int_0^{\bar{L}} \mathrm{d}t \,T(t)  =
 \int_{\tau_{\rm hydro}}^{L + \tau_{\rm hydro}} \mathrm{d}t \,T_{\mathrm{hydro}}(t) \,,
 \label{eq:0moment_powerlaw_HTL}
 \eeq
 \beq
\int_0^{\bar{L}} \mathrm{d}t \, t\,T(t)\,m_{\rm D}^2(t)  =
 \int_{\tau_{\rm hydro}}^{L + \tau_{\rm hydro}} {\rm d}t \, t\,T_{\rm hydro}(t)\, m^2_{{\rm D,\,hydro}}(t)\,,
\label{eq:1moment_powerlaw_HTL}
 \eeq
 \beq
 \int_0^{\bar{L}} \mathrm{d}t \, t^2\,T(t)\,m_{\rm D}^2(t)  
 = \int_{\tau_{\rm hydro}}^{L + \tau_{\rm hydro}} \mathrm{d}t \,t^2\,T_{\mathrm{hydro}}(t)\,m^2_{\mathrm{D,\, hydro}}(t) \,,
 \label{eq:2moment_powerlaw_HTL}
 \eeq
where \eqref{eq:1moment_powerlaw_HTL} ensures that the high-$\omega$ tail of the power-law spectrum matches that of the spectrum computed along the hydrodynamic path, as described in section~\ref{subsec:high-energy-tail}.

\begin{figure}
\includegraphics[width=\textwidth]{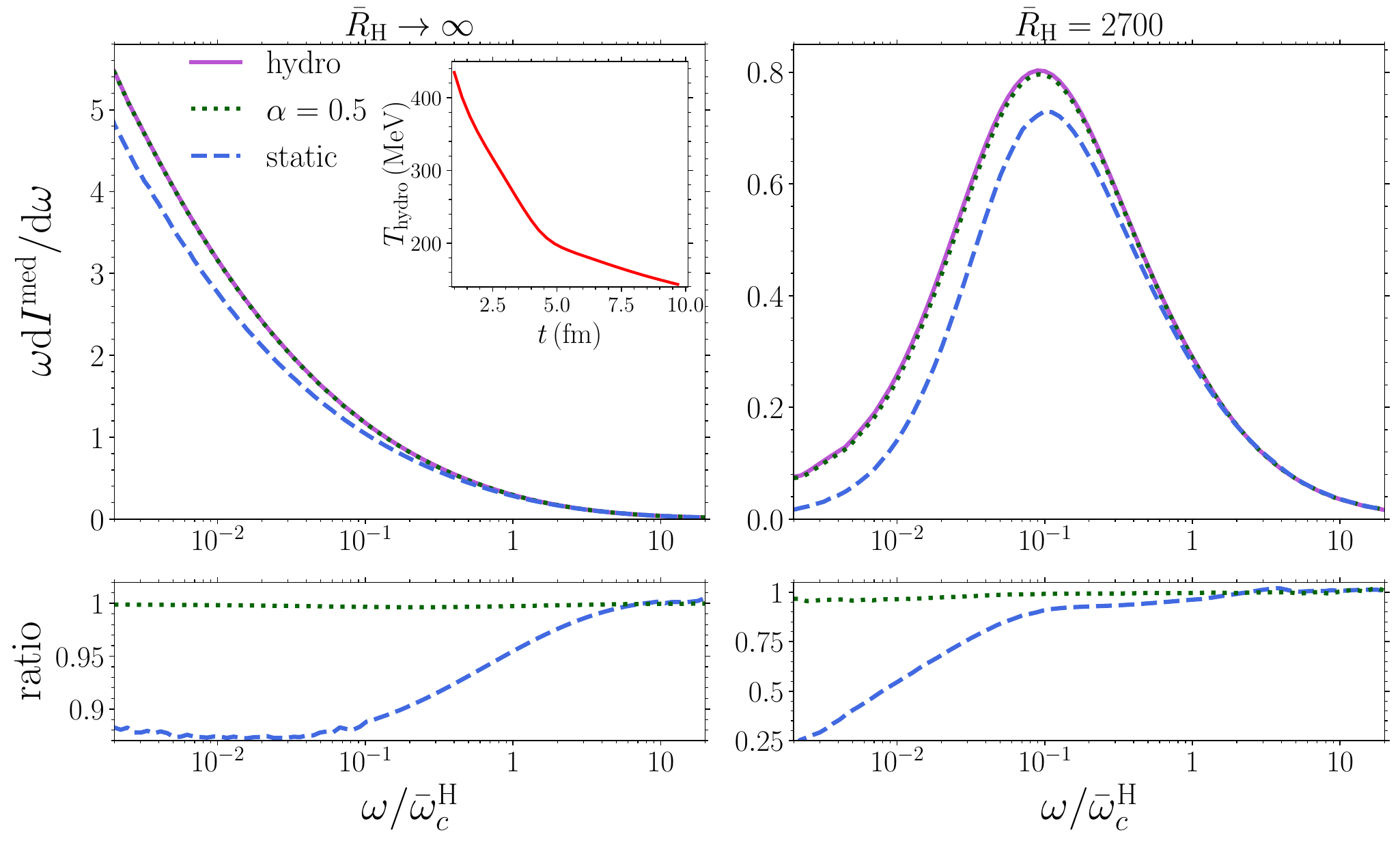}
\caption{
Top: fully resummed medium-induced gluon energy distribution for the HTL interaction model with $\chi_{\rm H} = 2$, and $ \bar R_{\rm H} \rightarrow \infty$ (left panel) or $ \bar R_{\rm H}  = 2700$ (right panel) as a function of $\omega/\bar \omega_c^{\rm H}$. The purple curves correspond to the spectra along the temperature profile shown in the inset figure sampled with a central production point over the 0-10$\%$ centrality class in $\sqrt{s_{\rm NN}} =5.02$ TeV Pb-Pb collisions at the LHC. For detailed information about the employed path, see main text. The dashed blue and dotted green curves correspond, respectively, to the results for the static (see eqs.~\eqref{eq:0moment_static_HTL}-\eqref{eq:2moment_static_HTL}) and power-law scenarios (see eqs.~\eqref{eq:0moment_powerlaw_HTL}-\eqref{eq:2moment_powerlaw_HTL}). Bottom: ratio of the power-law (static) spectrum w.r.t. the spectrum along the temperature profile in dotted green (dashed blue).}
\label{fig:HTL_LHC2_b3.2_x0y0theta0_a05}
\end{figure}

We present in figure~\ref{fig:HTL_LHC2_b3.2_x0y0theta0_a05} the gluon energy distribution computed along the same trajectory used in figure~\ref{fig:LHC2_b3.2_x0y0theta0_a05}: a straight-line path with a central production point sampled over the 0-10$\%$ centrality class in $\sqrt{s_{\rm NN}} =5.02$ TeV Pb-Pb collisions at the LHC, corresponding to a hard parton produced at the midpoint between the centers of the two lead nuclei and moving along the in-plane direction. The temperature profile is shown in the inset panel. We have set the parameters $c_1=0.20$ and $c_2=121$, which result for this path in $\chi_{\rm H}=2$ and $\bar R_{\rm H} \rightarrow \infty$ (left panel), and $\chi_{\rm H}=2$ and $R_{\rm H} = 2700$ (right panel). We compare this result to the static scenario obtained through the scaling laws \eqref{eq:0moment_static_HTL}-\eqref{eq:2moment_static_HTL} (dashed blue curve) and the proposed power-law scaling given by \eqref{eq:0moment_powerlaw_HTL}-\eqref{eq:2moment_powerlaw_HTL} (dotted green curve). We note that the energy distributions are plotted as a function of $\omega/\bar \omega_c^{\rm H}$, where $\bar \omega_c^{\rm H}$ is determined through~\eqref{eq:1moment_static_HTL}. As found for the Yukawa interaction model in figure~\ref{fig:LHC2_b3.2_x0y0theta0_a05}, the power-law scaling yields an accurate description of the spectrum along the actual path for all gluon energies, regardless of whether the kinematic constraint is removed (left panel) or imposed (right panel), while the static scaling fails to describe the spectrum along the hydrodynamic path for intermediate and low $\omega$'s.

\begin{figure}
\includegraphics[width=\textwidth]{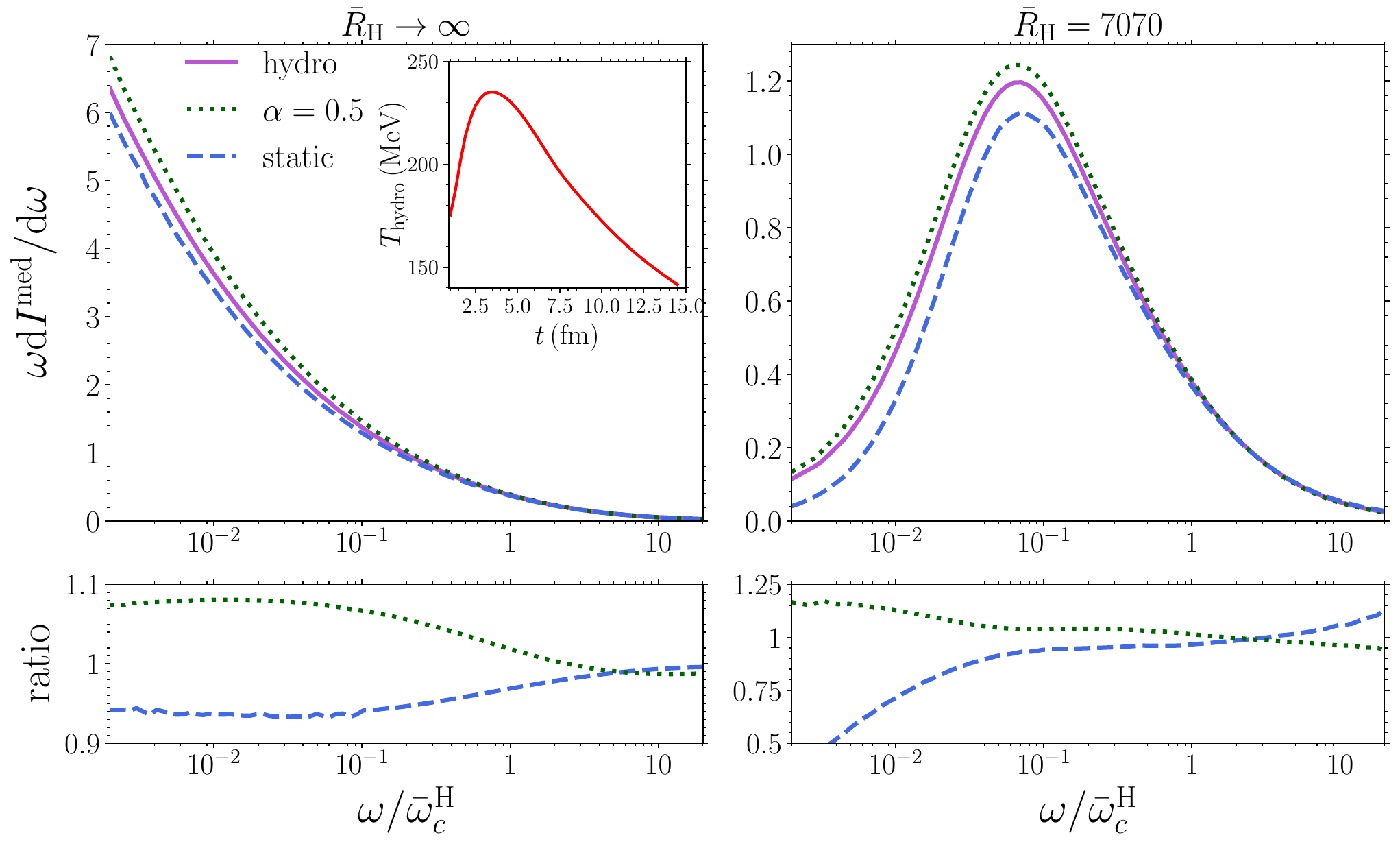}
\caption{ Top: fully resummed medium-induced energy distribution for the HTL interaction model with  $\chi_{\rm H}=2.7$, and $ \bar R_{\rm H}  \rightarrow \infty$ (left panel) or $ \bar R_{\rm H}  = 7070$ (right panel) as a function of $\omega/\bar{\omega}_c^{\rm H}$. The purple curves correspond to the spectra along the rare temperature profile shown in the inset figure sampled with a non-central production point over the 0-10$\%$ centrality class in $\sqrt{s_{\rm NN}} =5.02$ TeV Pb-Pb collisions at the LHC. For detailed information about the employed path, see main text. The dashed blue and dotted green curves correspond, respectively, to the spectra for the static (see eqs.~\eqref{eq:0moment_static_HTL}-\eqref{eq:2moment_static_HTL}) and power-law scenarios (see eqs.~\eqref{eq:0moment_powerlaw_HTL}-\eqref{eq:2moment_powerlaw_HTL}). Bottom: ratio of the power-law (static) spectrum w.r.t. the spectrum along the temperature profile in dotted green (dashed blue).}
\label{fig:HTL_LHC2_b3.2_x0y80theta255}
\end{figure}

We illustrate in figure~\ref{fig:HTL_LHC2_b3.2_x0y80theta255} the performance of the power-law and static scalings for a trajectory whose temperature profile does not monotonically decrease with time (as shown in the inset panel). For this purpose, we make use of the path that depicts the most extreme scenario in terms of temperature increase at initial times, as done for the Yukawa interaction model in  figure~\ref{fig:LHC2_b3.2_x0y80theta255}. This path corresponds to a hard parton produced at $\tau_{\rm hydro}$ 8\,fm away in the out-of-plane direction from the midpoint between the centers of the nuclei, and propagating inwards with a 255$^\circ$ angle with respect to the in-plane direction. In this figure, we set the parameter values $c_1=0.2$  and $c_2=121$  to the same values employed in figure~\ref{fig:HTL_LHC2_b3.2_x0y0theta0_a05}, which result for this path in  $\chi_{\rm H}=2.7$, and $\bar R_{\rm H} \rightarrow \infty$ (left panel), and $\chi_{\rm H}=2.7$ and $\bar R_{\rm H} =7070$ (right panel). We find that the power-law result deviates from the spectrum along the path for gluon energies below the characteristic gluon energy $\bar\omega_c^{\rm H}$. However, these deviations remain below $15\%$ across the entire range of gluon energies, being significantly smaller than the deviations observed in the static matching case. Therefore, we can confidently state that, even in this extreme scenario, the power-law scaling provides a more accurate description of the spectrum along the path compared to the static matching. Importantly, these results align perfectly with the findings obtained for the Yukawa parton-medium interaction model (see figure~\ref{fig:LHC2_b3.2_x0y80theta255} and its corresponding discussion). This consistency reinforces the robustness of the power-law scaling approach in capturing the behavior of the in-medium gluon spectrum, regardless of the specific parton-medium interaction model.

\bibliographystyle{JHEP}
\bibliography{references-2}

\end{document}